\documentclass[%
 aip,
 amsmath,amssymb,
 reprint,%
]{revtex4-1}

\usepackage{graphicx}
\usepackage{dcolumn}
\usepackage{bm}

\usepackage[utf8]{inputenc}
\usepackage[T1]{fontenc}
\usepackage{mathptmx}
\usepackage{etoolbox}
\usepackage{subcaption}
\usepackage{hyperref}
\usepackage{comment}
\usepackage{svg}
\usepackage{caption, xcolor}

\makeatletter
\def\@email#1#2{%
 \endgroup
 \patchcmd{\titleblock@produce}
  {\frontmatter@RRAPformat}
  {\frontmatter@RRAPformat{\produce@RRAP{*#1\href{mailto:#2}{#2}}}\frontmatter@RRAPformat}
  {}{}
}%
\makeatother

\begin{document}

\preprint{JAP}

\title[MuGrid-v2]{MuGrid-v2: A novel scintillator detector for multidisciplinary applications}
\author{Tao Yu}
\author{Yunsong Ning}%
\author{Yi Yuan}%
\author{Shihan Zhao}%
\author{Songran Qi}%
\author{Minchen Sun}%
\author{Yuye Li}%
\author{Zhirui Liu}%
\author{Aiyu Bai}
\author{Hesheng Liu}%
\affiliation{ 
  School of Physics, Sun Yat-sen University, 510275 Guangzhou, China \\
  Platform for Muon Science and Technology, Sun Yat-sen University, Guangzhou, China
  }%
\author{Yibo Lin}  
\author{Geng Tuo}
\author{Ting On Chan}
\affiliation{%
School of Geography and Planning, Sun Yat-sen University
}%
\author{Zhou Zhou}%
\author{Yu Chen}%
\email{chenyu73@mail.sysu.edu.cn}
\author{Jian Tang}
\email{tangjian5@mail.sysu.edu.cn}
\affiliation{ 
  School of Physics, Sun Yat-sen University, 510275 Guangzhou, China \\
  Platform for Muon Science and Technology, Sun Yat-sen University, Guangzhou, China 
  }%

\date{\today}

\begin{abstract}
Muography, traditionally recognized as a potent instrument for imaging the internal structure of gigantic objects, has initialized various interdisciplinary applications. 
As the financial and labor costs of muography detector development hinder their massive applications, we develop a novel muon detector called MuGrid by coupling a monolithic plastic scintillator with the light guide array in order to achieve competitive spatial resolution while substantially reducing production costs.
For a prototype detector in 30 cm $\times$ 30 cm, the intrinsic spatial resolution has been optimized toward a millimeter scale. 
An outdoor field muography experiment was conducted to monitor two buildings for validation purposes. The test successfully resolved the geometric influence of architectural features based on the attenuation of muon flux in a good agreement between experimental results and the simulation prediction. 
\end{abstract}

\maketitle


\section{\label{sec:1}Introduction}
As part of secondary products generated through primary cosmic-ray interactions with the atmosphere, cosmic muons exhibit exceptional penetrating capability owing to their weak electromagnetic interaction with matter. 
The majority of atmospheric muons propagate at relativistic velocities and maintain approximately straight trajectories~\cite{Tanaka2023} so that they are used to reconstruct the internal structure of objects in a non-deconstructive way. 
Muography with cosmic ray muons can be grouped into two categories: absorption muography~\cite{George:1955bzp} and scattering muography~\cite{Borozdin2003}.
The former is particularly suited for probing the internal structures of large-scale targets~\cite{Checchia2016}, while the latter is more applicable to detect smaller volumes of high Z materials through multiple Coulomb scattering analysis~\cite{Bonechi2020}. In this study, we focus on the absorption muography. 

Since the pioneering muography practice conducted by E.P. George in the 1950s~\cite{George:1955bzp}, this technique has evolved significantly and expanded into diverse fields, including geology~\cite{Teixeira2024,Beni2023,Nishiyama2014}, volcanology~\cite{Nagamine:1995np,Tanaka2009}, and archaeology~\cite{Alvarez1970,Liu2023}. 
The investigation of the hidden chamber within the Khafre's Pyramid in 2017~\cite{Morishima2017} brought great attention to muography, and subsequently researchers began to explore the potential of this technology in more fields.
In recent years, novel applications have emerged across multiple domains. 
For instance, H.K.M. Tanaka in Japan explored the potential possibility and experimentally demonstrated the feasibility of applying muography to atmospheric sciences~\cite{Tanaka2022a} and tide monitoring~\cite{Tanaka2021a,Tanaka2022}. Their team also proposed the idea of using muography for navigation and conducted demonstration experiments~\cite{Tanaka2020,Varga2024a,Chilingarian2024}.
Furthermore, applications in fields such as architectural science~\cite{Thompson2020}, blast furnace inspection~\cite{Cohu2023}, nuclear security~\cite{Borozdin2012,Procureur2023}, and even medical imaging~\cite{Morris2025} have been progressively proposed.

Detectors utilized in muography technology include nuclear emulsion detectors~\cite{Nakamura2006}, gas detectors~\cite{Gnanvo2011,jan2024}, and scintillator detectors~\cite{NiculescuOglinzanu2024,ning2025developmentportablecosmicraymuon}.
Both the first two categories can achieve very high spatial resolution up to several tens of $\mu\text{m}$ or even higher~\cite{Morishima2015,Nishio2015,Feng2021,Derre2001}. 
Although the spatial resolution of plastic scintillator detectors is difficult to keep up with other detector types, plastic scintillator detectors have gained increasing preference due to their robust environmental adaptability and ease of deployment in the emerging applications mentioned above.

Although plastic scintillators are relatively inexpensive compared to other detector types, achieving high spatial resolution often requires fabricating the scintillator into fine grained strips~~\cite{Bugg2014}, which substantially increases the expenses of material and processing~\cite{He2024}.
Furthermore, to achieve 3D muographic imaging, several measurements are required from different angles~\cite{Valencia:2024dtn}. Using multiple detectors can significantly reduce measurement durations and also mitigate the temporal impact from the seasonal modulation of muon flux~\cite{ice2019}, thus reducing systematic errors during data acquisition.
This further increased the cost of the entire experiment.
Although muography technology has been successfully implemented in numerous application domains, its manufacturing and deployment costs remain a persistent constraint on its further expansion into broader operational scenarios~\cite{Tanaka2022d}. 

To address this challenge, this study focuses on developing a detector system based on a monolithic plastic scintillator, which could substantially reduce the manufacturing costs of plastic scintillator processing. It also aims to achieve a spatial resolution of 1 cm or better, matching or even surpassing the performance of conventional strip scintillator detectors. This goal is accomplished through the use of light guide and wavelength shifting fibers.

We present the MuGrid-v2 detector design in Sec.~\ref{sec:design}, the readout electronic systems in Sec.~\ref{sec:electronics}, evaluation of the detector in Sec.~\ref{sec:performance}. Finally, in Sec.~\ref{sec:case}, a case study on campus is introduced.
\section{\label{sec:design}MuGrid detector Design}
\begin{figure*}[htbp]
  \centering
  \begin{subfigure}[b]{0.47\textwidth}
      \centering
      \includegraphics[height=7.5cm]{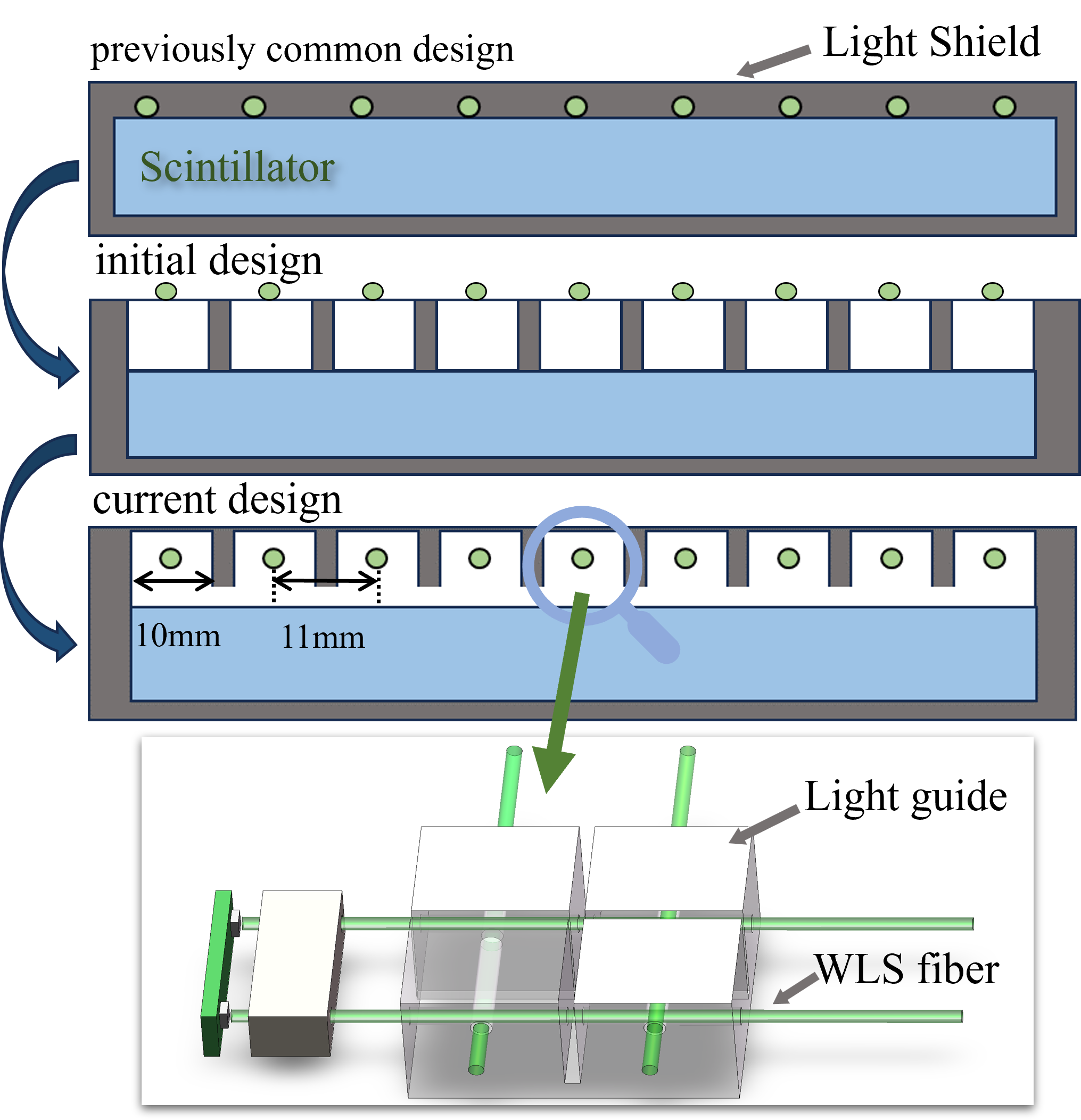}
      \caption{An evolution of the detector design.}
      \label{fig:airsubfig1}
  \end{subfigure}
  \hspace{0.6cm}
  \begin{subfigure}[b]{0.45\textwidth}
      \centering
      \includegraphics[height=7.3cm]{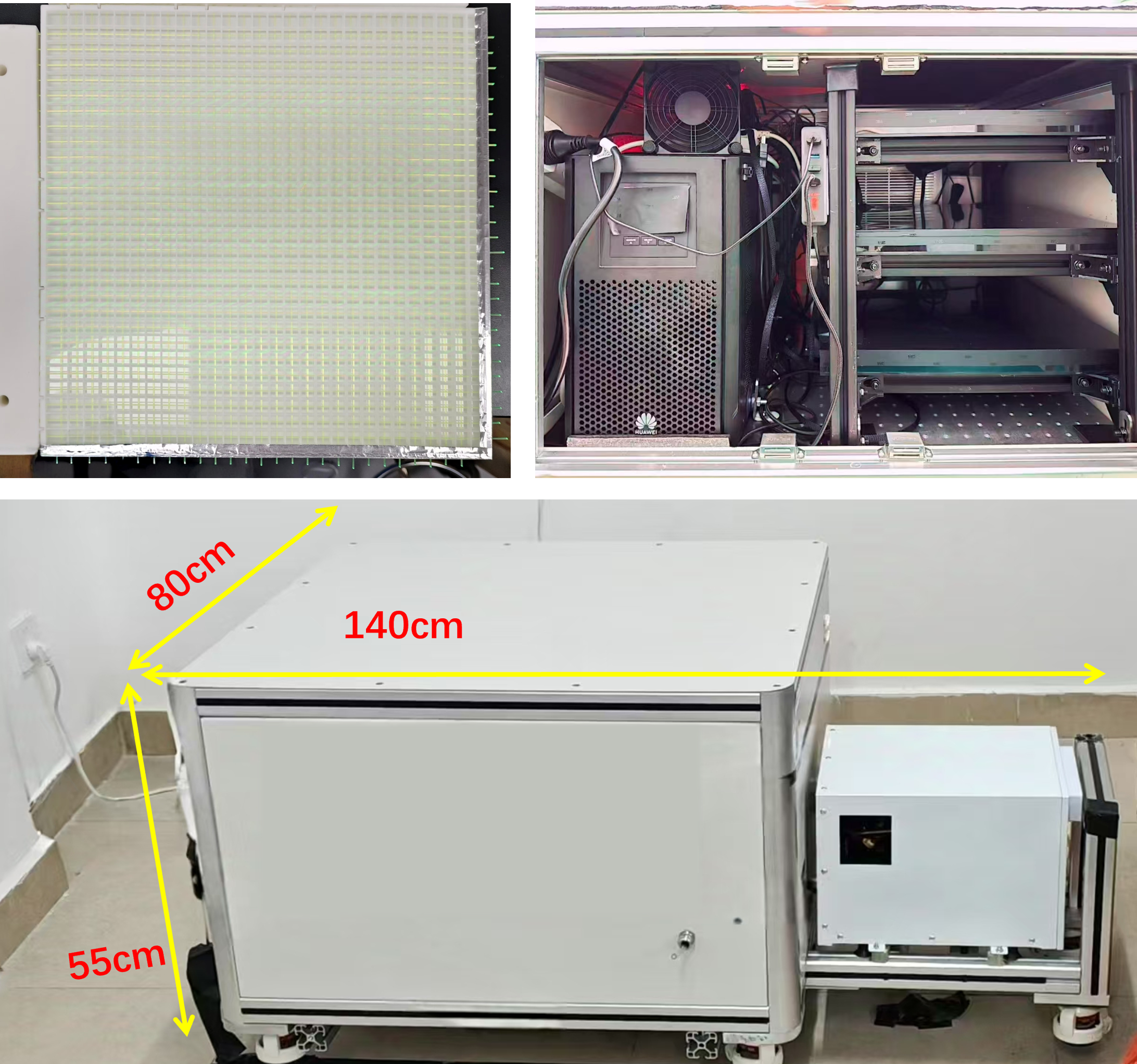}
      \caption{The detector MuGrid-v2 after assembly.}
      \label{fig:airsubfig2}
  \end{subfigure}
  \caption{The left panel illustrates the design evolution and concept of the MuGrid-v2 system. Compared to placing optical fibers on the light guide cubes, embedding them in the light guide structure provides a more stable fixation solution while also enhancing fiber collection efficiency for scintillation photons. The implementation of 3D printing technology makes the production of this complex structure relatively easy. The right panel shows the detector in assembly and the finished prototype.}
  \label{fig:v2desgin}
\end{figure*}
In the previous work~\cite{yang2022}, we proposed the concept of MuGrid-v1, which employed small light guide cubes to enhance the reconstruction accuracy for cosmic-ray muon signals as shown in Fig.~\ref{fig:v2desgin}. Compared to conventional approaches where optical fibers are directly coupled to plastic scintillators, this design enhances signal significance through the implementation of segmented light guides.

Subsequent design optimizations replace individual optical guide units with 3D printed optical grid arrays, significantly reducing both the installation procedure and associated costs. 
We first build a prototype based on the upgraded design and conduct preliminary tests. 
The test results demonstrated the diluted performance in a comparison with the simulation results. 
To figure out the reason, we visualize the raw SiPM data and examines the signal quality. Channels exceeding the threshold are marked with small triangles on the axis. 
 
\begin{figure}[htbp]
  \centering
  \begin{subfigure}[b]{0.40\textwidth}
      \centering
      \includegraphics[width=5.5cm]{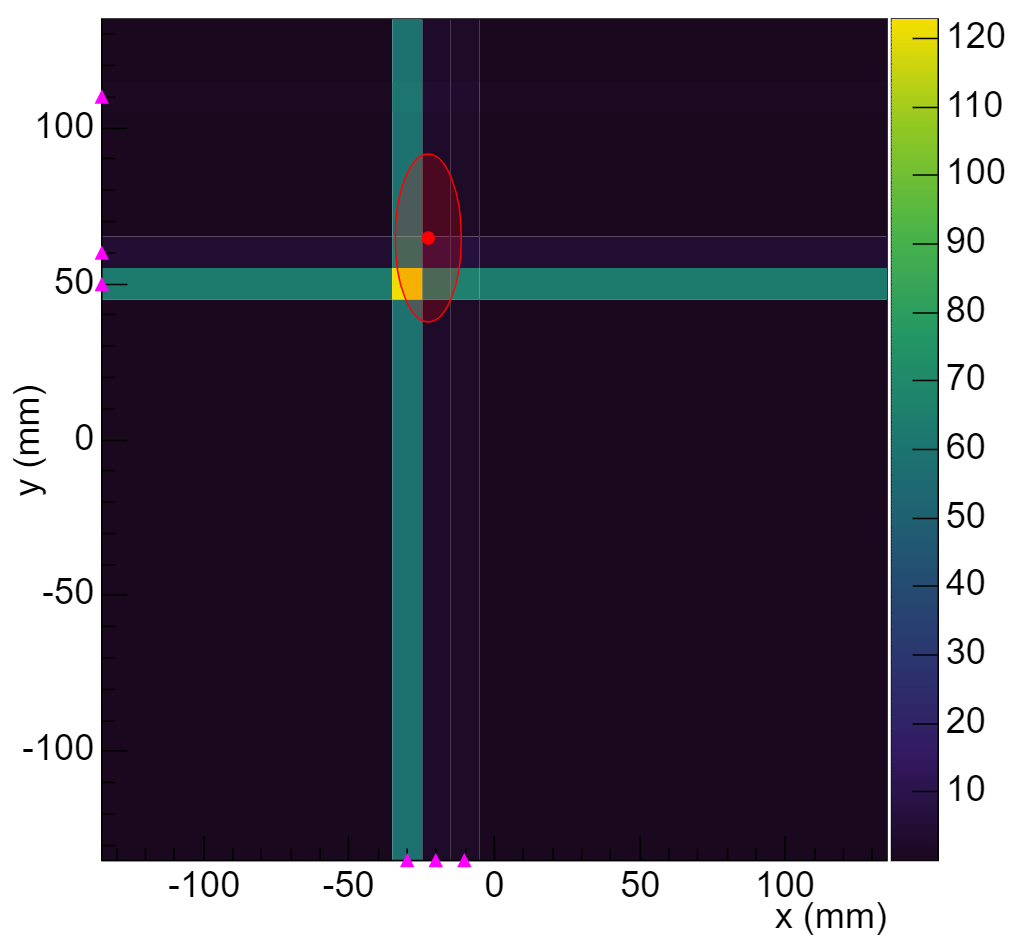}
      \caption{Typical signal in the original design.}
      \label{fig:v2ori}
  \end{subfigure}
  \hfill
  \begin{subfigure}[b]{0.40\textwidth}
      \centering
      \includegraphics[width=5.5cm]{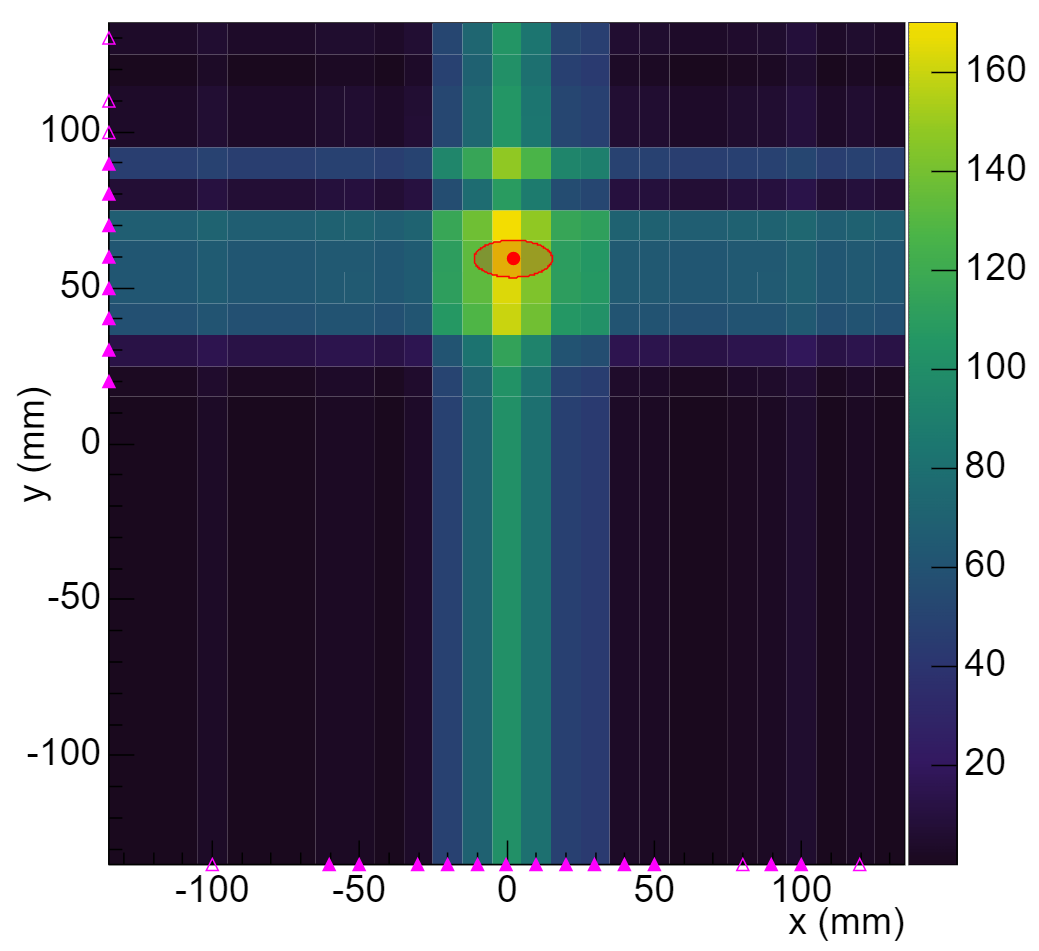}
      \caption{Typical signal after optimization. }
      \label{fig:v2air}
  \end{subfigure}
  \caption{The color of each channel corresponds to the signal amplitude. The red dot denotes the reconstructed position. }
  \label{fig:gridVis}
\end{figure}
The typical response of a muon hit from this prototype can be seen in Fig.~\ref{fig:v2ori}: only a limited number of channels are fired.
Through our investigation, we identified that the diluted performance is due to the strong attenuation of blue light emitted by the scintillator in the 3D-printed resin we used. This also leads to a quite low detection efficiency of the prototype (less than 20\%), which is a critical problem for muon detections.
\begin{figure*}[htbp]
\includegraphics[height = 8cm]{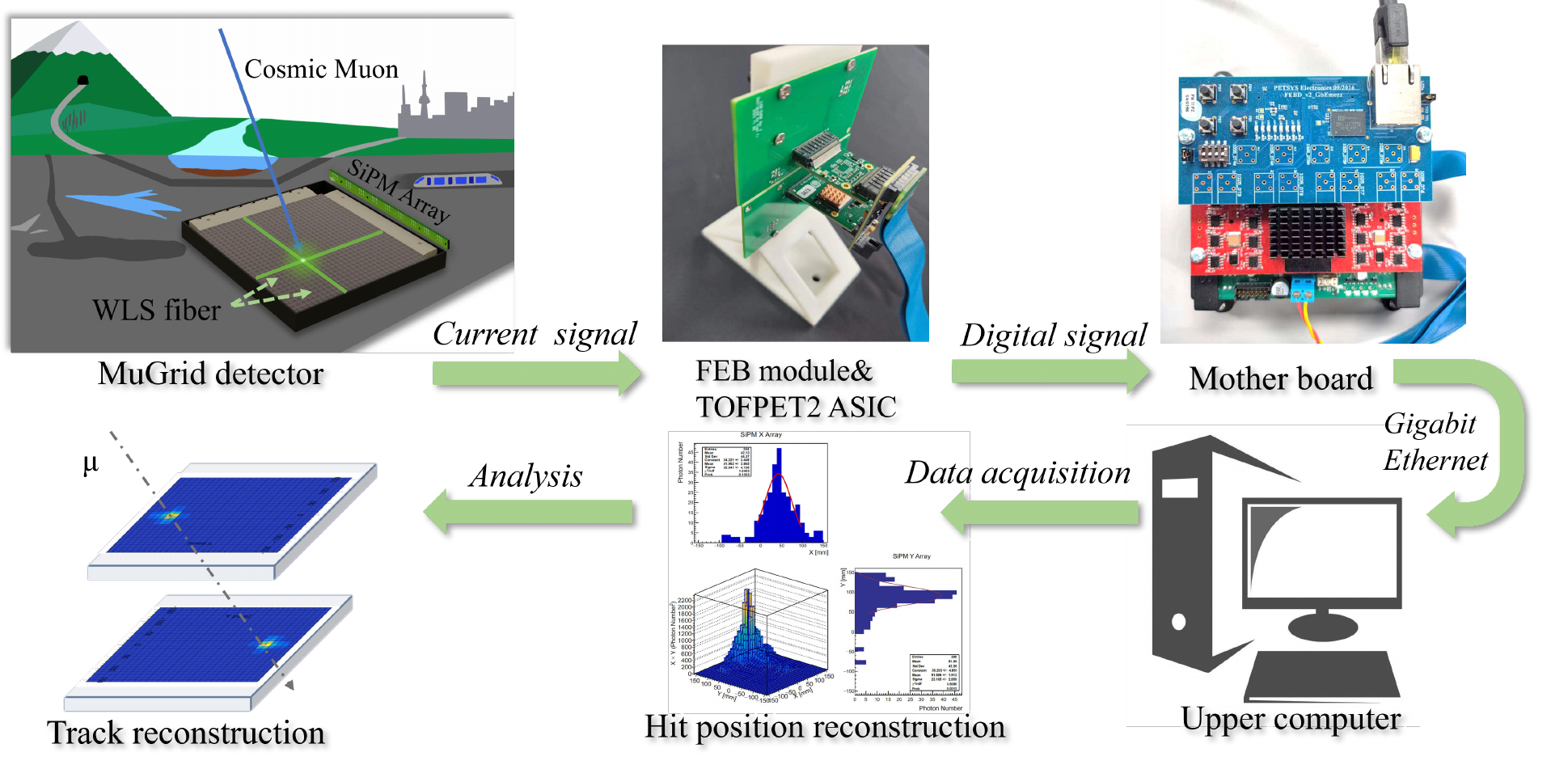}
\caption{\label{fig:flow}The overall workflow of the MuGrid detector.}
\end{figure*}
Moreover, an insufficient number of trigger channels increases the uncertainty in signal reconstructions, which is also shown in Fig.~\ref{fig:gridVis}.
To mitigate this issue, we perform both simulations and experimental work for a proper treatment of the scintillator surface. Ultimately, the bottom surface of the scintillator is modified with a diffuse reflective coating to enhance the number of fired channels. The segmentation method is replaced with high-reflectivity resin slices, while an enhanced specular reflector (ESR) film is applied to the upper layer to further improve its reflectivity.

The experimental results demonstrate that this configuration not only improves detection efficiency, but also reduces reconstruction errors, which could also be seen in Fig.~\ref{fig:gridVis}. 
An ellipse is used to represent the magnitude of the error, with its dimensions determined by the reconstruction errors in the x and y directions. Note that the ellipses are intentionally magnified to ensure visibility, as the uncertainty may be too small to be discernible in their original scale.

Subsequently, a detector prototype was manufactured, with a plastic scintillator module at a size of 30 cm $\times$ 30 cm $\times$ 1 cm~\cite{sp101}. The scintillator is segmented into 27 channels using a 3D printed light guide grid. The dimensions of the light guide remain as shown in Fig.\ref{fig:airsubfig1}. Wavelength shifting fibers (WLS) are embedded within the grid to collect photons emitted by the scintillator. The WLS fibers employed in the prototype are Kurrary's Y-11(200) fibers, with a diameter of 1 mm~\cite{kuraraypsfY11}. 
The S13360-1350 MPPCs from Hamamastu~\cite{hamamatsus13360}, are chosen as optical sensors, since their size is similar to the diameter of the WLS fiber. 
They also possess the advantages of high gains, excellent time resolution, robustness, and being cost-effective enough~\cite{Piemonte2019}. 
The current signals generated by 27 SiPMs in each row are transmitted to the SiPM adapter board through a micro-coaxial ribbon cable. 
Subsequent signal processing is taken by the TOFPET2 ASIC evaluation kit~\cite{Francesco2016}, in which the front-end board (FEB/S) is replaced by our self-designed adapter board mentioned before.  
Each chip can provide 64 individual readout channels with a maximum hit rate of 480 kcps per channel. After an analog-to-digital conversion, the motherboard in the evaluation kit will transmit the data to the upstream data acquisition software through Ethernet. 
The raw data are stored in ROOT format files, which contain information such as hit channel ID, hit time, charge integration result (QDC) and integration time (ToT). 
We can see the workflow of the MuGrid-v2 detector in Fig.~\ref{fig:flow}.

\section{\label{sec:electronics}front-end readout electronics}
In this work, we adapt the TOFPET2 ASIC evaluation kit as front-end readout electronics to boost the development cycle.
The motherboard provided in the kit supports readout up to 1024 channels. Each channel features an independent dynamic range of 1500 pC and a TDC binning resolution of 30 ps.
Detailed performance evaluation of the TOFPET2 ASIC can be found in reference~\cite{Bugalho2019}.
The following section focuses on the parameter optimization process we undertake to tailor the TOFPET system to our specific application requirements.

The TOFPET2 ASIC is equipped with independent amplifiers, discriminators, time-to-digital converters (TDCs) and charge-to-digital converters (QDCs) for each channel. 
The input current signal will be replicated into 3 branches: T, E, and Q.
Details of its logic are also shown in the literature~\cite{Bugalho2019}.

As SiPMs exhibit higher noise levels compared to photomultiplier tubes (PMTs), we use the TOFPET2 ASIC evaluation kit to test different settings of operating parameters in order to obtain an appropriate combination.

TOFPET2 ASIC provides two measurement modes: Time and Charge (QDC) mode and Dual time (ToT) mode.
QDC mode, which is more commonly used, has a quite large dynamic range up to 4000 photoelectrons, since the TOFPET2 ASIC is originally designed for positron emission tomography (PET) applications. 
The inorganic scintillation crystals used in the PET system exhibit significantly higher light yields compared to plastic scintillators, requiring the dynamic range of the chip to be large enough to cover the 511 keV positron annihilation signals encountered in PET. 
This feature of TOFPET2 poses challenges for the precise discrimination of low-amplitude signals such as a few tens of photoelectrons.
To the best of our knowledge, no prior studies have utilized the TOFPET2 ASIC to calibrate the multi-photoelectron spectrum of SiPMs.
Since a reconstruction of the muon hit position relies on the accurate signal amplitude, it is essential to discriminate photoelectron numbers in the relatively precise way for the MuGrid-v2 detector.
Considering the superior time resolution of the TOFPET2 ASIC, we choose the ToT mode to measure the SiPM signal. 

After determining the measurement mode, a suitable trigger logic needs to be selected as the TOFPET2 ASIC is equipped with 3 distinct discriminators: $D\_E$, $D\_T_1$, and $D\_T_2$.
We evaluated the trigger mode provided by TOFPET2 ASIC, including the single threshold trigger mode, the dual threshold trigger mode, and others.
Since timing jitter between different discriminators may cause negative ToT values which is observed in the evaluation, the single threshold trigger with only $D\_T_1$ discriminators is chosen as the trigger mode.

\begin{figure}[htbp]
  \includegraphics[width=0.45\textwidth]{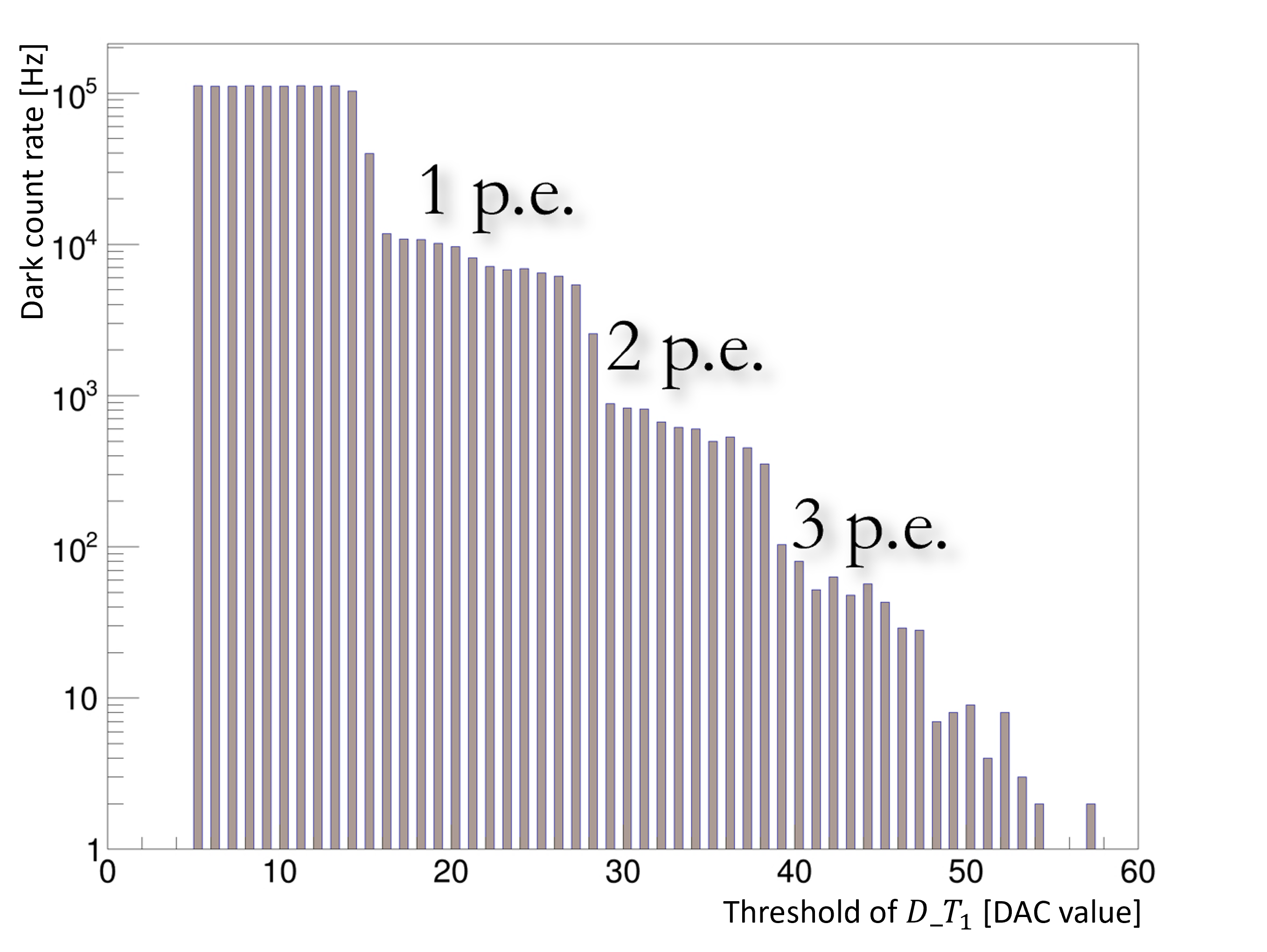}
  \caption{\label{fig:DCR}The variation of dark count rates with an increase of the discriminator threshold. At each step, the number of photoelectrons points to a specific threshold plateau.}
\end{figure}

After an establishment that the combination of ToT mode and single $D\_T_1$ achieves a small signal identification, the next step is to determine the appropriate discriminator threshold. To reach the goal, a dark count test is performed by placing the SiPM in a light-tight dark chamber without coupling any scintillator to the photosensor. 
Fig~\ref{fig:DCR} shows the dark count rate as a function of the discriminator threshold. 
A distinct "step-like" effect (a sharp drop in the count rate) can be observed. When the threshold exceeds the amplitude, it corresponds to a specific photoelectron level.
In this technique, we can get the average amplitude of single-photon pulses.
Based on the measured curve, we set the $D\_T_1$ threshold value of 32 (in DAC units), corresponding to about 2.5 p.e.
This threshold is high enough to reject most of noise pulses while preserving the detection efficiency.

\begin{figure}[htbp]
    \centering
    \begin{subfigure}[b]{0.44\textwidth}
        \centering
        \includegraphics[width=8cm]{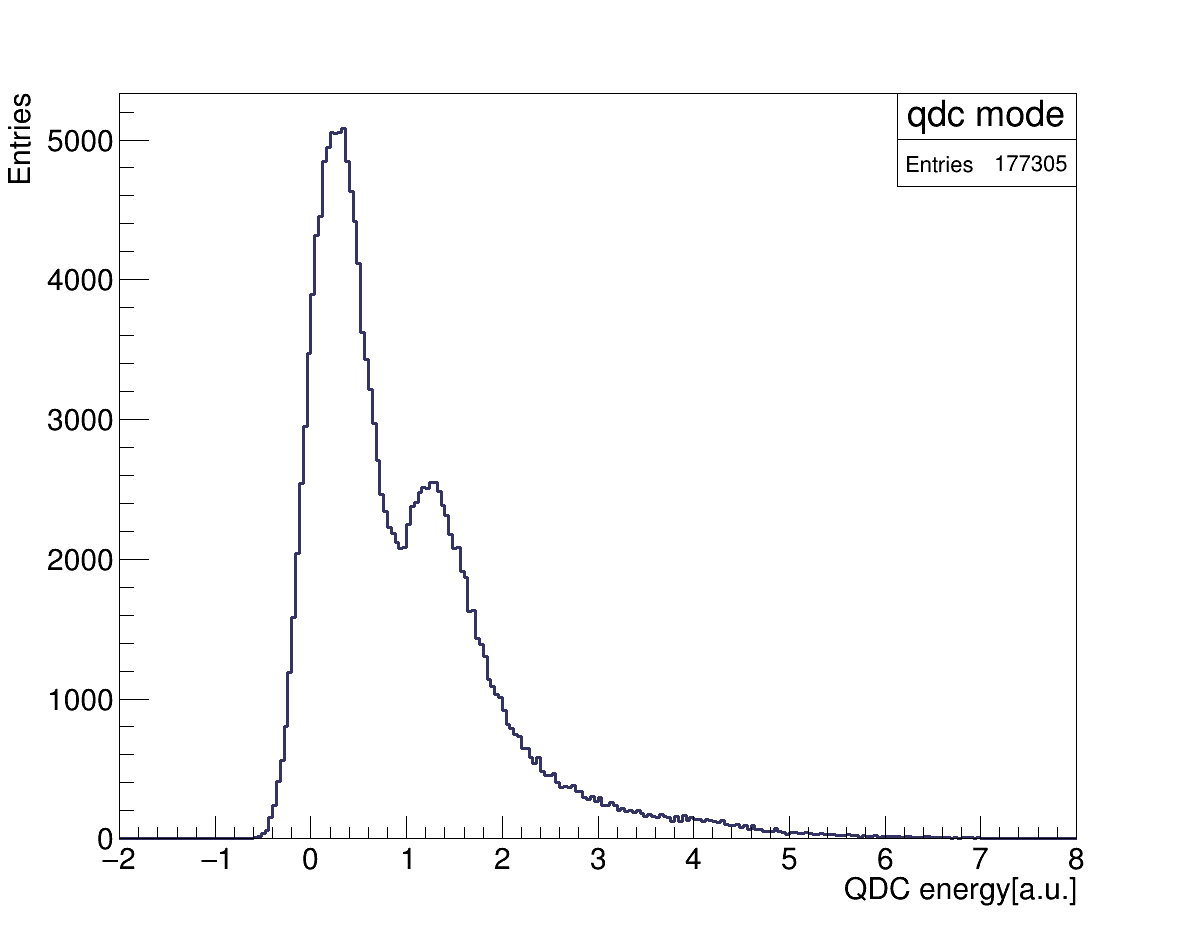}
        \caption{qdc mode results}
        \label{fig:multipesubfig1}
    \end{subfigure}
    \begin{subfigure}[b]{0.44\textwidth}
        \centering
        \includegraphics[width=8cm]{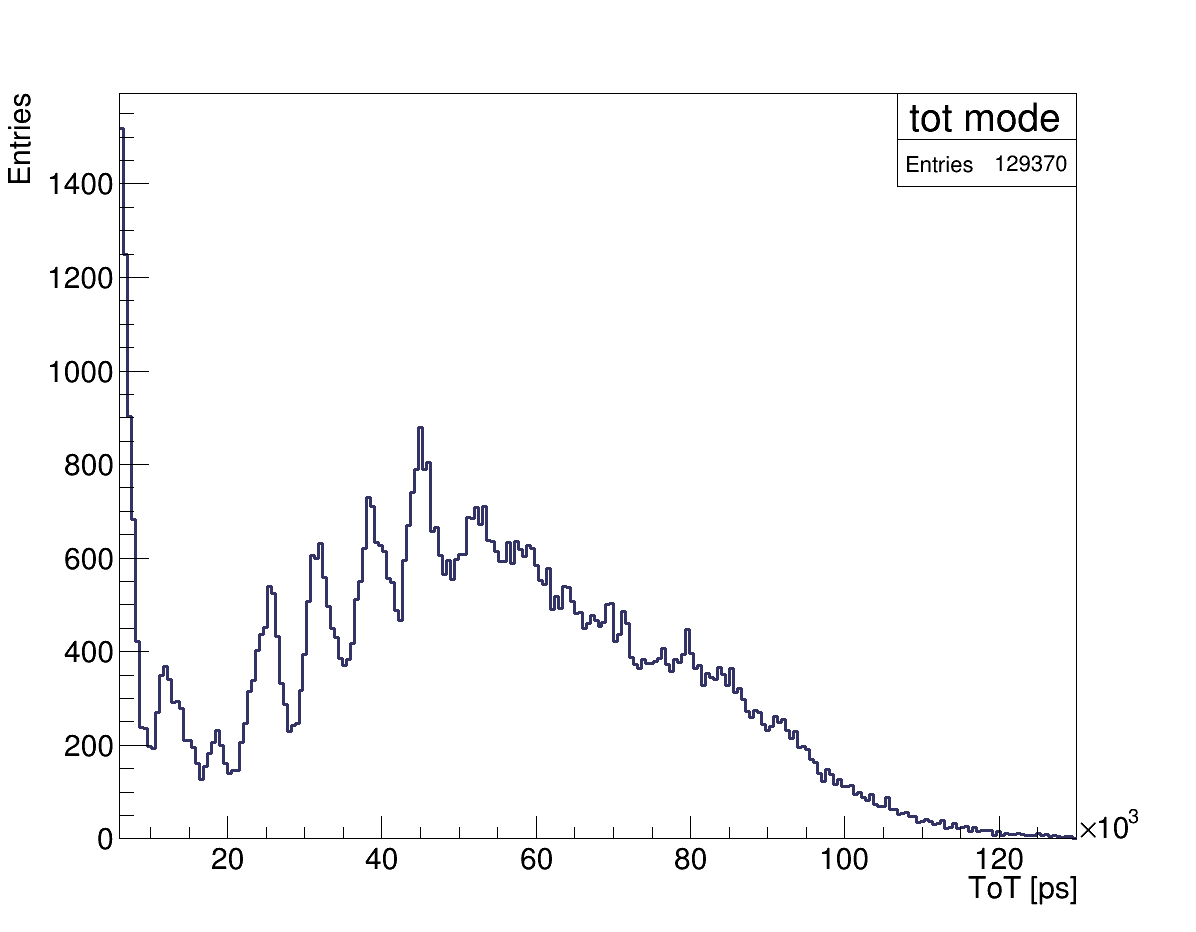}
        \caption{ToT mode results }
        \label{fig:multipesubfig2}
    \end{subfigure}
    \caption{multi-photoelectron spectrum measured with the S13360-1350 biased at 4.5 V overvoltage in both QDC and ToT mode. It is worth noting that a dip appears in both figure. This splitting feature is not caused by different  photoelectron peaks but rather results from signal reflection induced by impedance mismatch at the ASIC input interface.}
    \label{fig:multipe}
\end{figure}

With this configuration, we perform a multi-photoelectron spectrum test for the S13360-1350 SiPMs.
As illustrated in Fig.~\ref{fig:multipe}, a well-resolved multi-photoelectron spectrum is clearly visible. 
In comparison, the top panel of Fig.~\ref{fig:multipe} presents the results obtained by the QDC mode in the same configuration, which has almost no capability of identifying photoelectron numbers.

All measurements above are performed after a recalibration process, in which the current noise level is evaluated to optimize the system parameters adaptively. 
In practical applications, the threshold can be dynamically adjusted according to the specific noise condition of the operating environment. Note that it is not necessary to recalibrate the system unless the ASIC parameters are modified.

\section{\label{sec:performance}performance of MuGrid detector}

\subsection{Detection efficiency}
Detection efficiency is a critical parameter for the detector, as it determines the data-taking time required for actual muography applications.
Following the strategy in the reference~\cite{Grupen2023}, we measure the detection efficiency of the MuGrid detector. 
we establish a three-layer muon telescope to measure the detection efficiency.
The distance between two layers is set to 15 cm. 
The coincidence timing window is set to 15$\text{ns}$.
The efficiency is calculated as the ratio of events that trigger all three layers to those of the upper and lower layers triggered only.
By exchanging the order of three layers, we can get the efficiency of each layer.
For each layer, we collected 95356, 93658, and 153365 valid events respectively.
The test results show that the detection efficiencies of three detector layers in our prototype are 86.12$\pm$ 0.09\%, 90.83$\pm$ 0.11\%, and 87.72 $\pm$ 0.08\%.

\subsection{Spatial resolution calibration}
\begin{figure}[htbp]
  \includegraphics[width=0.42\textwidth]{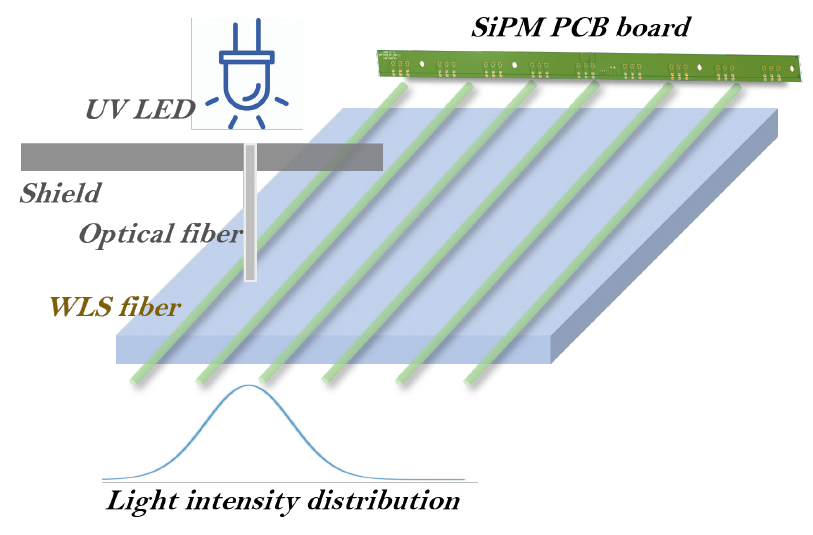}
  \caption{\label{fig:UVLED} The schematic diagram of the LED calibration experiment setup. The optical fiber receives LED light from a little hole to approximate a point light source.}
\end{figure}

The intrinsic spatial resolution of the detector is another key performance metric for muon detections. 
Two conventional methods are commonly employed to evaluate spatial resolution. One involves taking a higher resolution detector as a reference, such as a high-resolution gas detector, while the other utilizes a well-understood radioactive source or light source to calibrate the detector~\cite{DomingoPardo2009}.
Due to the laboratory constraints, we follow the second way and took a 380 nm ultraviolet (UV) LED as a light source to excite the scintillator. The simple setup is illustrated in Fig.~\ref{fig:UVLED}.

We select five test points on the detector plane with four points at each corner and one point at the center to perform the resolution calibration. Each point is illuminated by a $1\,\text{kHz}$ UV LED in 300 seconds to prepare enough mock data. 

The reconstructed interaction positions are indicated by red markers. We first applied a centroid reconstruction algorithm given by Eq.~\ref{eq:E} to mock the muon hit position:
\begin{equation}
  \label{eq:E}
  \hat{X} = \sum E_i X_i / \sum E_i
\end{equation}
where $X_i$ is the position of the $i$-th SiPM, and $E_i$ is the the number of photoelectrons recorded of the $i$-th SiPM.

\begin{figure*}[htbp]
  \centering
  \begin{subfigure}[b]{0.48\linewidth}
    \centering

    \includegraphics[height=8cm]{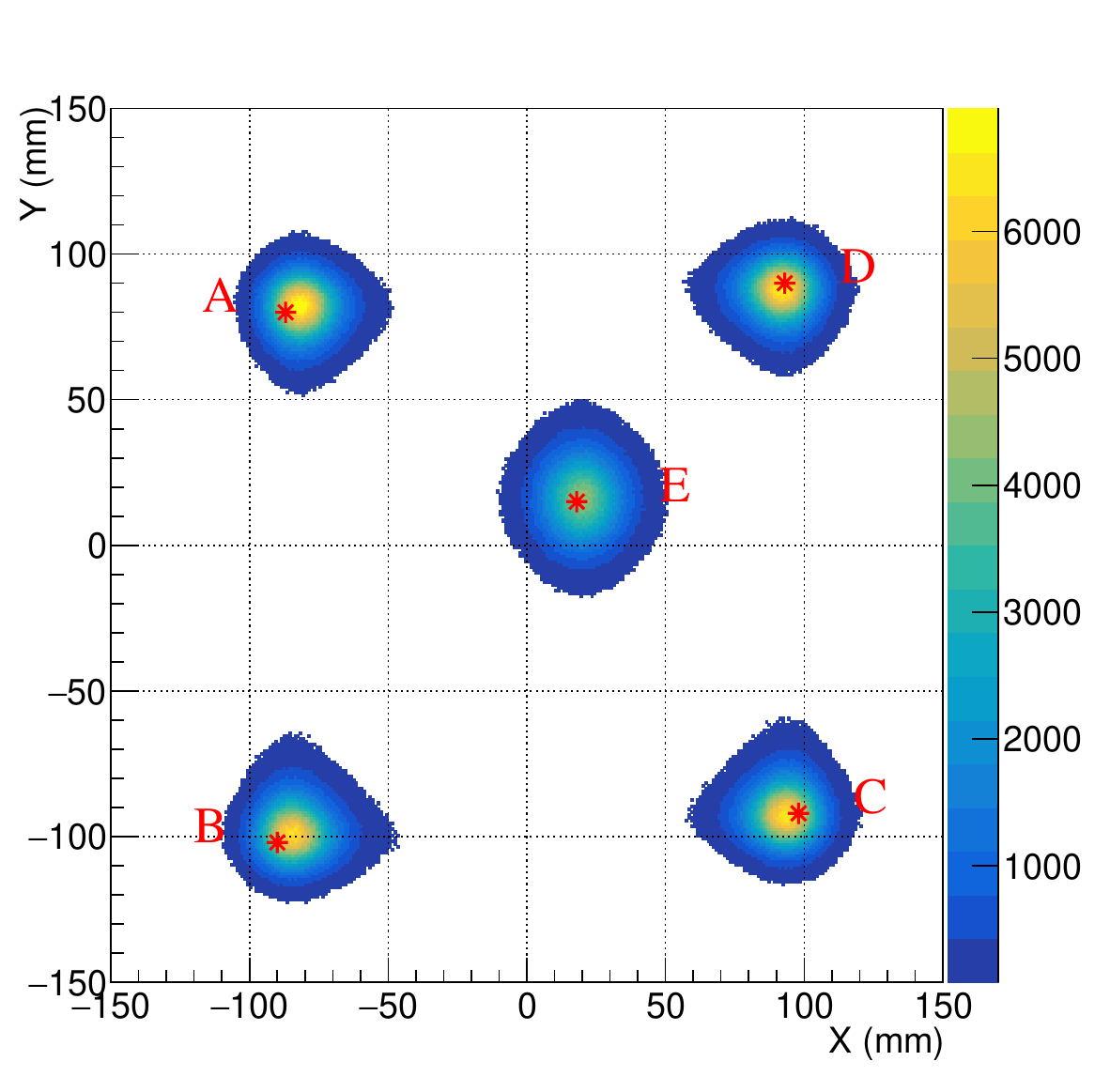}
    \caption{Calibrations with the energy only.}
    \label{fig:5pointsub1}
  \end{subfigure}
  \hfill
  \begin{subfigure}[b]{0.48\textwidth}
    \centering
    \includegraphics[height=8cm]{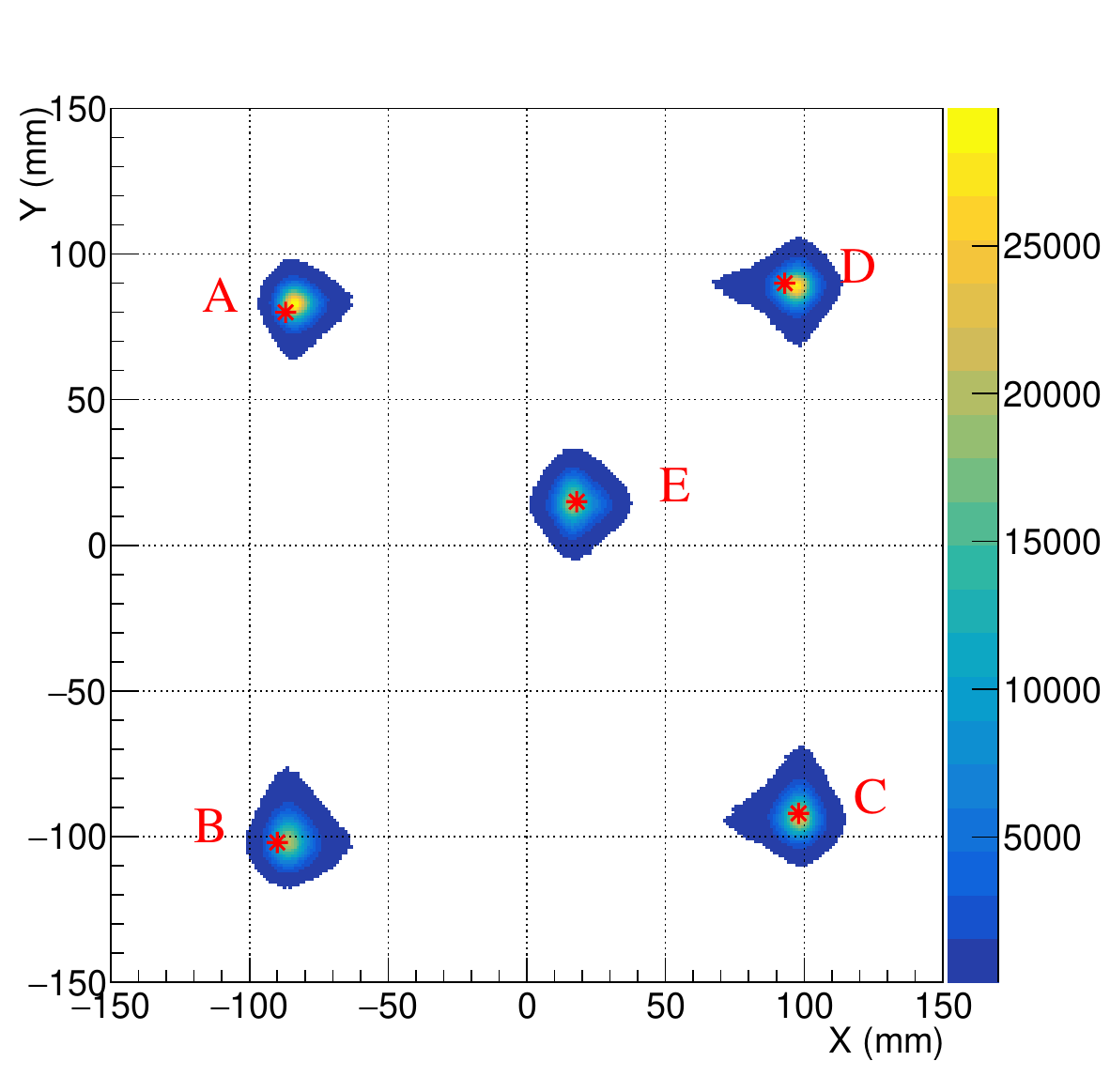}
    \caption{Calibrations with the energy and timestamp information.}
    \label{fig:5pointsub2}
  \end{subfigure}
  \caption{Reconstruction results. Each red asterisk indicates the actual position of the optical fiber signal. 
  }
  \label{fig:5point}
\end{figure*}

We present the results of the calibration experiment in Fig.~\ref{fig:5point}. We also perform a Gaussian fitting at each test point. A spatial resolution ($\sigma$) of approximately 9.2 mm was obtained by averaging the results of five points. In particular, the reconstructed positions near the corners exhibit systematic shifts toward the center. 
The deviation occurs because of the light yield difference caused by the detector geometry. The lower light yield can be collected by wavelength shifting fiber at the periphery than that in the central region. This will result in the diminished weight for edge positions and a consequent centroid bias.
    
\begin{figure}[htbp]
  \includegraphics[width=0.42\textwidth]{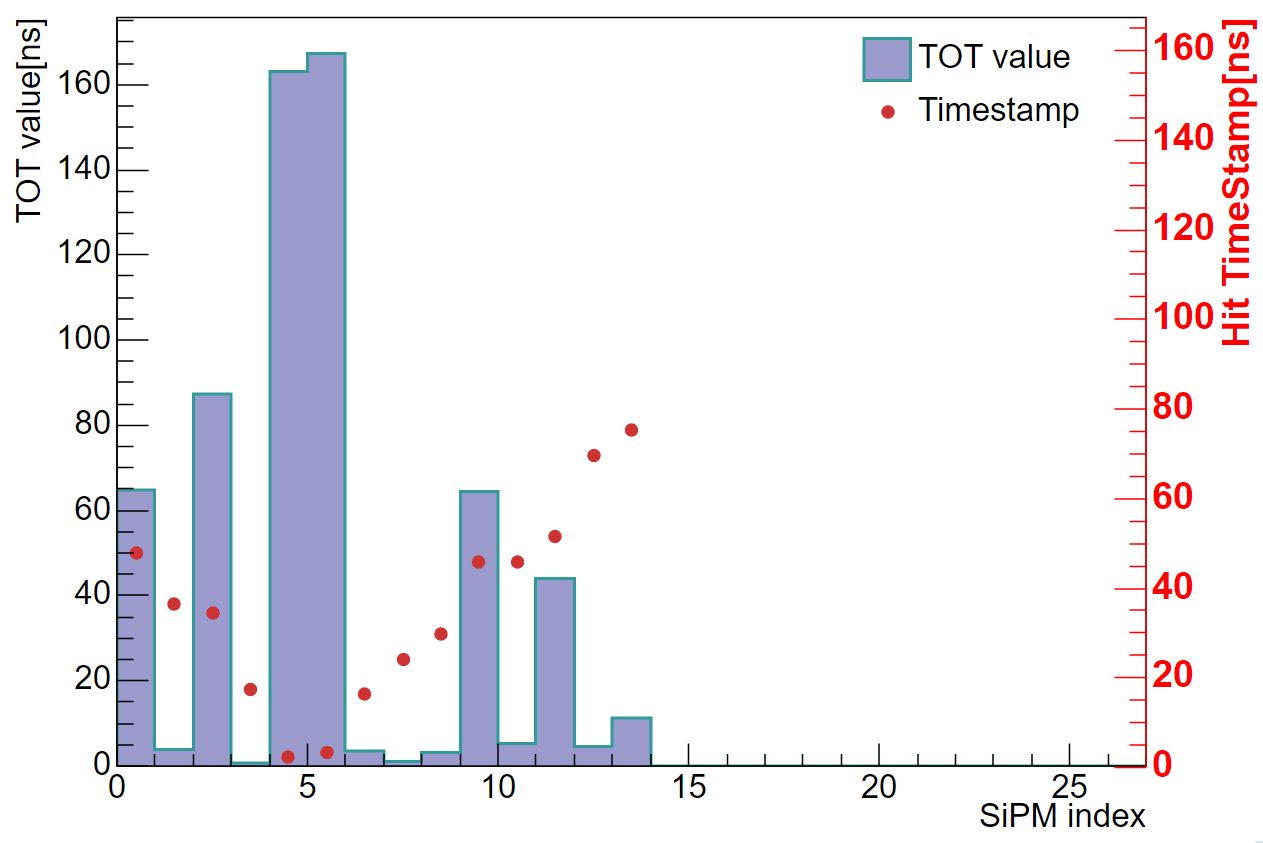}
  \caption{\label{fig:tandE} The histogram displays the Time-over-Threshold (ToT) values for trigger channels on one side of our detector. The red markers indicate the hit timestamp of each channel, with the timestamp of the first triggered channel is defined as zero time reference.}
\end{figure}

To improve the fidelity of reconstruction, we investigate additional information available in the raw data. 
In the ToT mode of TOFPET2, the QDC and ToT branches record totally identical data but with units of nanoseconds (ns) and picoseconds (ps) respectively. 
Consequently, the useful parameter is only the hit time stamp. 
By correlating the hit time with the signal amplitude in Fig.~\ref{fig:tandE}, we observe that channels closer to the interaction point tend to register earlier SiPM signals. Previous studies~\cite{Cossio2018} report a similar phenomenon.

It indicates that time information can also be used to assist in event reconstructions, particularly in scenarios where the distribution of energy information is not ideal, as is shown in Fig.~\ref{fig:tandE}.
We then introduced a time-based weight into the reconstruction algorithm, with the following weighting:
\begin{equation}
  \label{eq:time}
  \hat{X} = \sum \frac{E_i X_i}{\sqrt{1+(t_i/\sigma_t)^2}} / \sum \frac{E_i}{\sqrt{1+(t_i/\sigma_t)^2}}
\end{equation}
where $t_i$ is the hit time of the $i$th SiPM, $\sigma_t$ is the standard deviation of timestamp in this event.

The reconstruction results obtained by this method are presented in Fig.~\ref{fig:5pointsub2}, where the reconstructed region exhibits a striking contraction compared to the centroid reconstruction method. The average resolution in this algorithm is 4.6 mm, which is much better than the centroid method.
Moreover, this algorithm reduces the weighting of channels located far from the center of interaction, partially mitigating the positional bias observed in events reconstructed by the centroid method.
\section{\label{sec:case}outdoor demo test}

\subsection{Field compatible chassis}
To support field experiments, we design a temperature-controlled enclosure that provides stable power supply to the detector system. The enclosure is equipped with an air conditioning unit to maintain the stable environment, as the performance of SiPM and ASIC is sensitive to the temperature. The chassis features only two external interfaces: a power supply port and an Ethernet port. 
A miniaturized router connected to the Ethernet port enables a network relay to the computer inside the closure. Thus, the detector supports remote real-time monitoring, operation and data retrieval in two network configurations, where both wired and wireless connections can be established. 
In addition, the system incorporates an uninterruptible power supply (UPS) to protect against equipment damage induced potentially by sudden power outages. 
All internal components are isolated with the shock absorption mechanism to mitigate vibration-induced disturbances during transportation. Since the total weight of the system is around 170 kg, three pairs of wheels are mounted on the base of the enclosure, facilitating an easy relocation of the detector system.
A photograph of the prototype enclosure and its internal structural have already presented in Fig.~\ref{fig:airsubfig2}

\subsection{Open-sky measurement}

A field test is conducted in an open area on the Sun Yat-sen University campus (23.194°N, 113.306°E).

\begin{figure}[htbp]
  \includegraphics[width=0.48\textwidth]{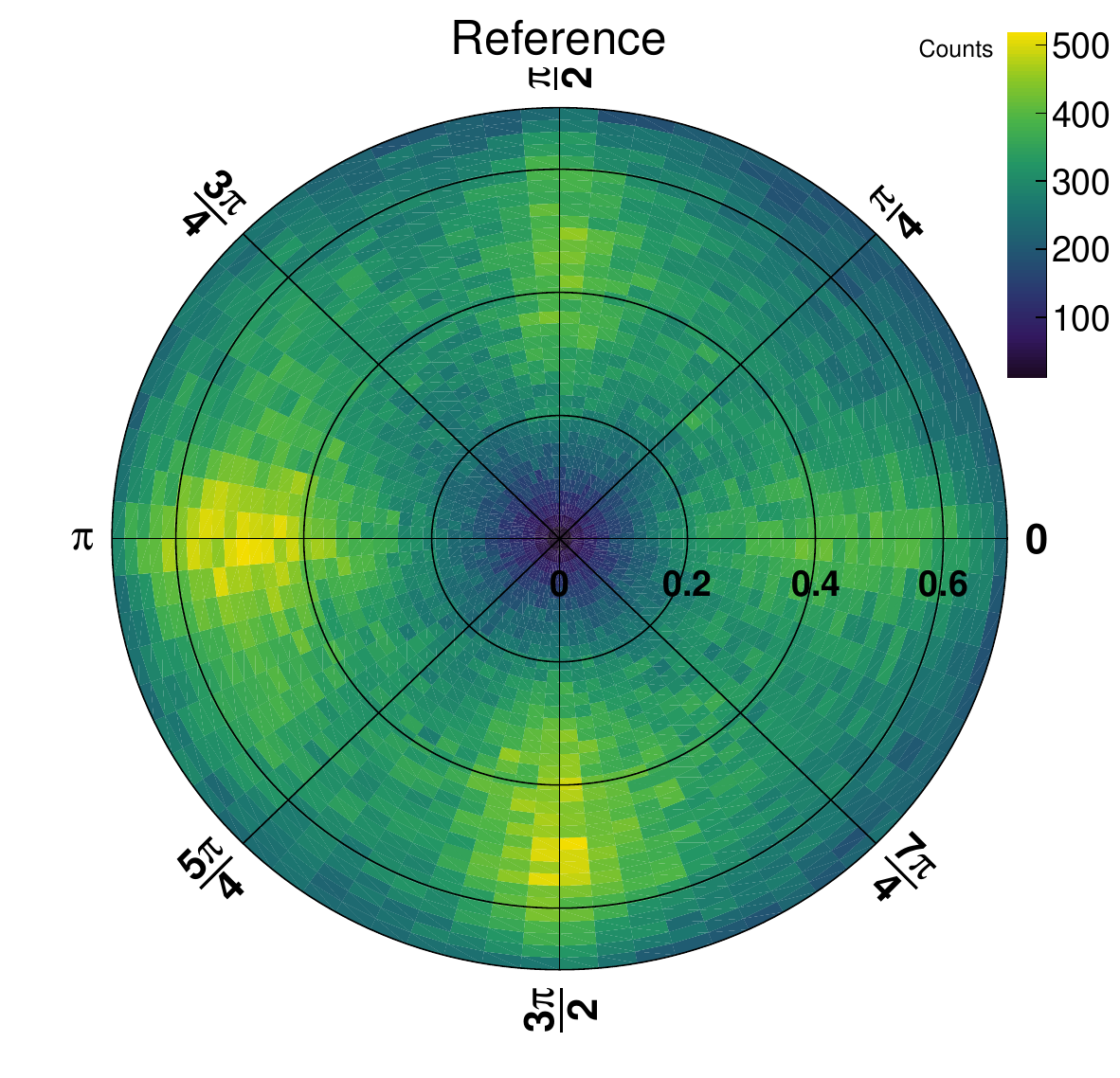}
  \caption{\label{fig:opensky} Results of the open-sky measurement. The excess of muon flux observed along the four directions is a geometric effect induced by the square-shape detector.}
\end{figure}

After calibrating the equipment, a 48-hour data acquisition session is conducted and a total of 769718 moun events were collected.
This dataset serves as a baseline for normalizing subsequent experiments.
The distribution of cosmic muon zenith angle measured is shown in Fig.~\ref{fig:opensky} in a polar coordinate system.

\begin{figure*}[ht!]
  \centering
  \begin{subfigure}[b]{0.36\linewidth}
    \centering
    \includegraphics[height=5.2cm]{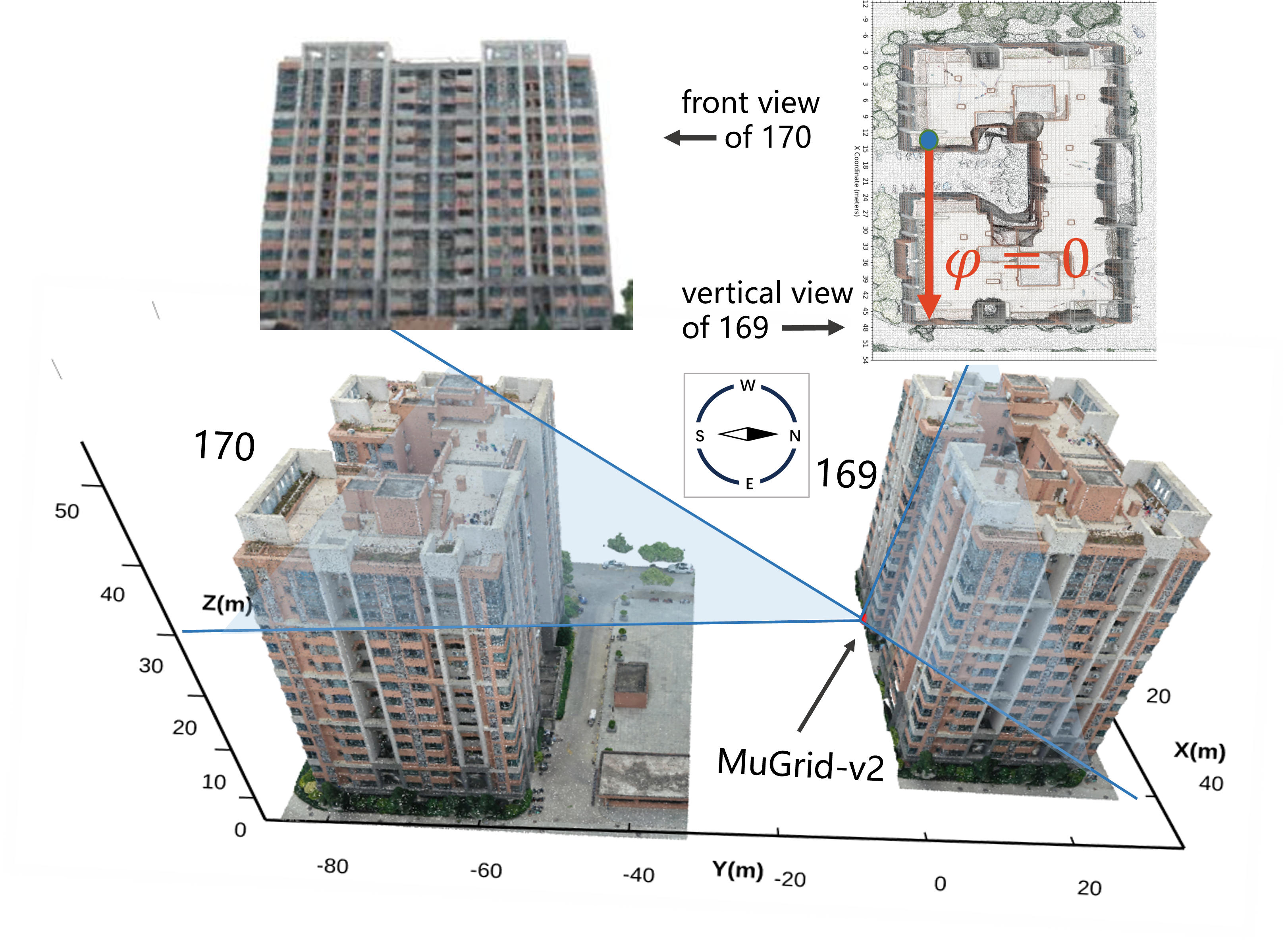}
    \caption{An illustration of experimental setup.}
    \label{fig:170setup}
  \end{subfigure}
  \begin{subfigure}[b]{0.31\textwidth}
    \centering
    \includegraphics[height=5.0cm]{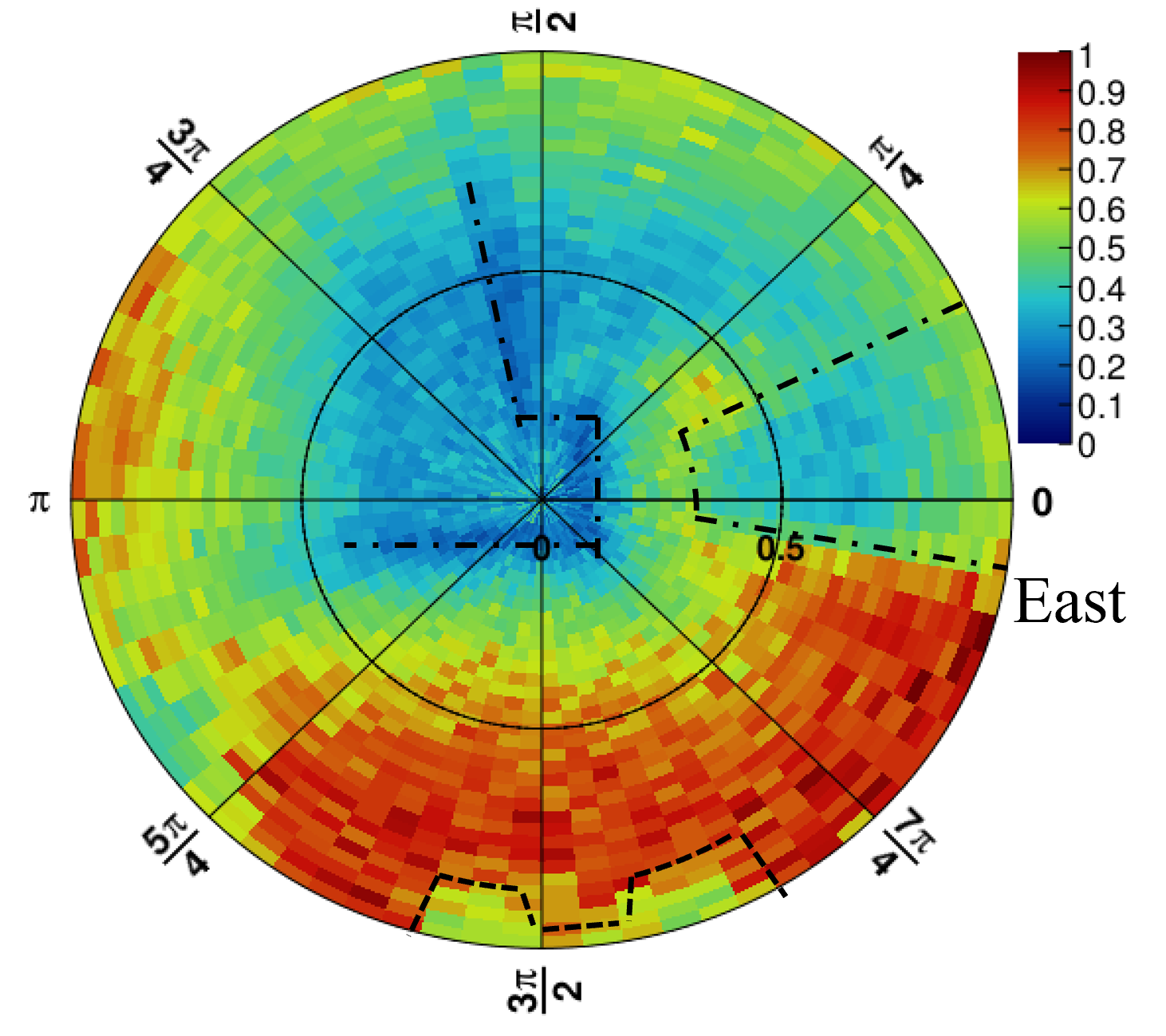}
    \caption{Muon survival ratio in simulation.}
    \label{fig:resultcompare_sim}
  \end{subfigure}
  \begin{subfigure}{0.31\textwidth}
          \centering
    \includegraphics[height=5.0cm]{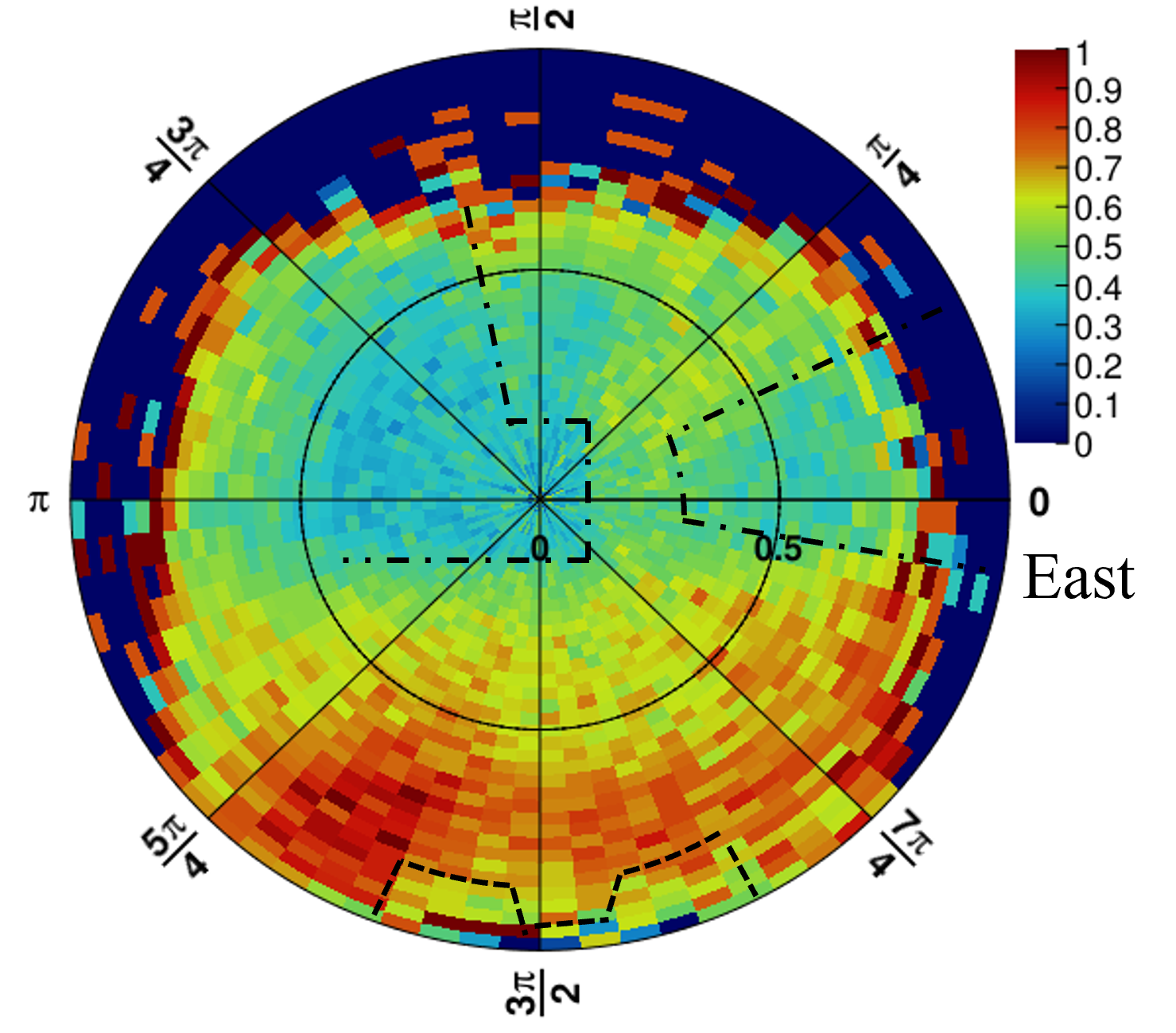}
    \caption{Muon survival ratio in experiment.}
    \label{fig:resultcompare_exp}
  \end{subfigure}
  \caption{(a): The red dot indicates the position of the detector. The building contours are represented by colorized point clouds. The blue semi-transparent surfaces roughly illustrate the range of incident muon rays reaching the detector. The two graphs (b) and (c) on the right side show the survival ratios obtained by simulation and experimental measurements, respectively.}
  \label{fig:result170}
\end{figure*}
\subsection{Case study: a building "shadow" by muography}
The MuGrid-v2 prototype is deployed in a student dormitory building at Sun Yat-sen University, and the imaging target is the shadow caused by the absorption of the muon by the two buildings labeled by 169 and 170, respectively. Using the structure from motion (SofM) technique with images captured by unmanned aerial vehicles, we obtain a three-dimensional point cloud dataset of the external profile of each building with millimeter-resolution and their spatial relationships. 
The point cloud data are later utilized as input for Monte Carlo simulations in the custom software. 
The experimental setup, including the geometry of the imaging target and the detector location, is depicted in Fig.~\ref{fig:170setup}.

To verify the experimental results, we build a digital elevation model simulation (DEM) using TURTLE~\cite{Niess:2019hdn}, based on data from aerial vehicles.
The simulated cosmic ray muon flux is generated by EcoMug~\cite{Zurlo:2022ltm}.
Then a Monte Carlo simulation is performed in the DEM building using PUMAS~\cite{Niess2022} and Geant4~\cite{Agostinelli2003} in order to generate reference data for the experiment. The spatial resolution of the detector used in the simulation is based on the results of our calibration result in Sec.\ref{sec:performance}. To simply account for the non-uniform density distribution of the building structure,  the exterior walls were assigned a density of $2.3\, \text{~g/cm}^3$ , while the internal structures were modeled with a reduced density of $0.3\, \text{~g/cm}^3$ in simulation~\cite{instruments6040077}.
For the sake of a quantitative analysis, the muon attenuation is converted to the survival ratio with the muon flux passing through the building divided by the reference muon flux in the open sky.
The detector deployed in Building 170 is supposed to identify the directional variations of muon flux induced by the U-shaped building geometry, as evidenced by the simulation results in Fig.~\ref{fig:resultcompare_sim}.
The blue regions illustrate the absorption effect caused by the two vertical sections of the building, while the red region represents the muon flux from the open sky, which is nearly unattenuated.
The yellow area at the bottom of the figure corresponds to Building 170 located on the north side.

Based on the experimental results containing 769,718 muon tracks,  Fig.~\ref{fig:resultcompare_exp} shows that we can effectively reconstruct the flux pattern caused by the geometry of the building. Given the limited data acquisition period of 96 hours, the collected data statistics remain sparse, particularly at low zenith angles. However, these results are sufficient to demonstrate the operational feasibility of the MuGrid-v2 detector.
In particular, an asymmetry is observed in the experimental flux distribution due to the $5^\circ$ southward tilt applied to the detector system during the data-taking period, which extends the detector’s field of view towards Building 170. 
In future implementations, we plan to expand the system to a grid structure, enabling a $2 \pi$ effective solid angle coverage comparable to the first generation of detector MuGrid-v1.

\section{Conclusion and outlook}
We have developed an innovative structural design of the muography detector to achieve high detection efficiency and spatial resolution on a monolithic plastic scintillator. 
Taking advantages of the exceptional temporal resolution of the system, we have integrated the timing information into the reconstruction process, which indeed improved precision.
The first prototype demonstrate an intrinsic spatial resolution of 4.6 mm, which is expected to improve with further optimization of the reconstruction algorithm.
An outdoor field test has been conducted, validating the stability and operational effectiveness of the detector prototype under non-laboratory environmental conditions.
We will soon deploy the detector system MuGrid-v2 for interdisciplinary studies, such as mines or traffic tunnels.
The detector geometry rendering and basic event reconstruction modules have already been implemented (as shown in the visualization plot Fig.~\ref{fig:result170}). Future iterations will prioritize usability enhancements, including graphical workflow builders and automated parameter optimization, to empower non-experts to conduct on-site simulation and data analysis for dedicated applications efficiently.

\begin{acknowledgments}
This project is supported by Fundamental Research Funds for the Central Universities (23xkjc017) in Sun Yat-sen University, Guangdong Basic and Applied Basic Research Foundation under Grant No. 2025A1515010669 and Natural Science Foundation of Guangzhou under Grant No. 2024A04J6243. The authors appreciate the strong support from SYSU Experimental Physics Center and School of Systems Science and Engineering for an access to the venue for open-sky measurement. We also appreciate the valuable email discussion with Dr. Ricardo from PETsys Electronics S.A.

\end{acknowledgments}

\section*{Data Availability Statement}
The data that support the findings of this study are available from the corresponding author upon reasonable request.

\section*{References}
\bibliography{aipsamp}

\providecommand{\noopsort}[1]{}\providecommand{\singleletter}[1]{#1}%
\begin{thebibliography}{53}%
\makeatletter
\providecommand \@ifxundefined [1]{%
 \@ifx{#1\undefined}
}%
\providecommand \@ifnum [1]{%
 \ifnum #1\expandafter \@firstoftwo
 \else \expandafter \@secondoftwo
 \fi
}%
\providecommand \@ifx [1]{%
 \ifx #1\expandafter \@firstoftwo
 \else \expandafter \@secondoftwo
 \fi
}%
\providecommand \natexlab [1]{#1}%
\providecommand \enquote  [1]{``#1''}%
\providecommand \bibnamefont  [1]{#1}%
\providecommand \bibfnamefont [1]{#1}%
\providecommand \citenamefont [1]{#1}%
\providecommand \href@noop [0]{\@secondoftwo}%
\providecommand \href [0]{\begingroup \@sanitize@url \@href}%
\providecommand \@href[1]{\@@startlink{#1}\@@href}%
\providecommand \@@href[1]{\endgroup#1\@@endlink}%
\providecommand \@sanitize@url [0]{\catcode `\\12\catcode `\$12\catcode `\&12\catcode `\#12\catcode `\^12\catcode `\_12\catcode `\%12\relax}%
\providecommand \@@startlink[1]{}%
\providecommand \@@endlink[0]{}%
\providecommand \url  [0]{\begingroup\@sanitize@url \@url }%
\providecommand \@url [1]{\endgroup\@href {#1}{\urlprefix }}%
\providecommand \urlprefix  [0]{URL }%
\providecommand \Eprint [0]{\href }%
\providecommand \doibase [0]{http://dx.doi.org/}%
\providecommand \selectlanguage [0]{\@gobble}%
\providecommand \bibinfo  [0]{\@secondoftwo}%
\providecommand \bibfield  [0]{\@secondoftwo}%
\providecommand \translation [1]{[#1]}%
\providecommand \BibitemOpen [0]{}%
\providecommand \bibitemStop [0]{}%
\providecommand \bibitemNoStop [0]{.\EOS\space}%
\providecommand \EOS [0]{\spacefactor3000\relax}%
\providecommand \BibitemShut  [1]{\csname bibitem#1\endcsname}%
\let\auto@bib@innerbib\@empty
\bibitem [{\citenamefont {Tanaka}\ \emph {et~al.}(2023)\citenamefont {Tanaka}, \citenamefont {Bozza}, \citenamefont {Bross}, \citenamefont {Cantoni} \emph {et~al.}}]{Tanaka2023}%
  \BibitemOpen
  \bibfield  {author} {\bibinfo {author} {\bibfnamefont {H.~K.~M.}\ \bibnamefont {Tanaka}}, \bibinfo {author} {\bibfnamefont {C.}~\bibnamefont {Bozza}}, \bibinfo {author} {\bibfnamefont {A.}~\bibnamefont {Bross}}, \bibinfo {author} {\bibfnamefont {E.}~\bibnamefont {Cantoni}},  \emph {et~al.},\ }\bibfield  {title} {\enquote {\bibinfo {title} {Muography},}\ }\href {\doibase 10.1038/s43586-023-00270-7} {\bibfield  {journal} {\bibinfo  {journal} {Nature Reviews Methods Primers}\ }\textbf {\bibinfo {volume} {3}},\ \bibinfo {pages} {1--17} (\bibinfo {year} {2023})}\BibitemShut {NoStop}%
\bibitem [{\citenamefont {George}(1955)}]{George:1955bzp}%
  \BibitemOpen
  \bibfield  {author} {\bibinfo {author} {\bibfnamefont {E.~P.}\ \bibnamefont {George}},\ }\bibfield  {title} {\enquote {\bibinfo {title} {{Cosmic Rays Measure Overburden of Tunnel}},}\ }\href@noop {} {\bibfield  {journal} {\bibinfo  {journal} {Commonwealth Engineer}\ }\textbf {\bibinfo {volume} {42}},\ \bibinfo {pages} {455--457} (\bibinfo {year} {1955})}\BibitemShut {NoStop}%
\bibitem [{\citenamefont {Borozdin}\ \emph {et~al.}(2003)\citenamefont {Borozdin}, \citenamefont {Hogan}, \citenamefont {Morris}, \citenamefont {Priedhorsky}, \citenamefont {Saunders}, \citenamefont {Schultz},\ and\ \citenamefont {Teasdale}}]{Borozdin2003}%
  \BibitemOpen
  \bibfield  {author} {\bibinfo {author} {\bibfnamefont {K.~N.}\ \bibnamefont {Borozdin}}, \bibinfo {author} {\bibfnamefont {G.~E.}\ \bibnamefont {Hogan}}, \bibinfo {author} {\bibfnamefont {C.}~\bibnamefont {Morris}}, \bibinfo {author} {\bibfnamefont {W.~C.}\ \bibnamefont {Priedhorsky}}, \bibinfo {author} {\bibfnamefont {A.}~\bibnamefont {Saunders}}, \bibinfo {author} {\bibfnamefont {L.~J.}\ \bibnamefont {Schultz}}, \ and\ \bibinfo {author} {\bibfnamefont {M.~E.}\ \bibnamefont {Teasdale}},\ }\bibfield  {title} {\enquote {\bibinfo {title} {Radiographic imaging with cosmic-ray muons},}\ }\href {\doibase 10.1038/422277a} {\bibfield  {journal} {\bibinfo  {journal} {Nature}\ }\textbf {\bibinfo {volume} {422}},\ \bibinfo {pages} {277--277} (\bibinfo {year} {2003})}\BibitemShut {NoStop}%
\bibitem [{\citenamefont {Checchia}(2016)}]{Checchia2016}%
  \BibitemOpen
  \bibfield  {author} {\bibinfo {author} {\bibfnamefont {P.}~\bibnamefont {Checchia}},\ }\bibfield  {title} {\enquote {\bibinfo {title} {Review of possible applications of cosmic muon tomography},}\ }\href {\doibase 10.1088/1748-0221/11/12/C12072} {\bibfield  {journal} {\bibinfo  {journal} {Journal of Instrumentation}\ }\textbf {\bibinfo {volume} {11}},\ \bibinfo {pages} {C12072} (\bibinfo {year} {2016})}\BibitemShut {NoStop}%
\bibitem [{\citenamefont {Bonechi}, \citenamefont {D'Alessandro},\ and\ \citenamefont {Giammanco}(2020)}]{Bonechi2020}%
  \BibitemOpen
  \bibfield  {author} {\bibinfo {author} {\bibfnamefont {L.}~\bibnamefont {Bonechi}}, \bibinfo {author} {\bibfnamefont {R.}~\bibnamefont {D'Alessandro}}, \ and\ \bibinfo {author} {\bibfnamefont {A.}~\bibnamefont {Giammanco}},\ }\bibfield  {title} {\enquote {\bibinfo {title} {{Atmospheric muons as an imaging tool}},}\ }\href {\doibase 10.1016/j.revip.2020.100038} {\bibfield  {journal} {\bibinfo  {journal} {Rev. Phys.}\ }\textbf {\bibinfo {volume} {5}},\ \bibinfo {pages} {100038} (\bibinfo {year} {2020})},\ \Eprint {http://arxiv.org/abs/1906.03934} {arXiv:1906.03934 [physics.ins-det]} \BibitemShut {NoStop}%
\bibitem [{\citenamefont {Teixeira}\ \emph {et~al.}(2024)\citenamefont {Teixeira}, \citenamefont {Blanco}, \citenamefont {Caldeira} \emph {et~al.}}]{Teixeira2024}%
  \BibitemOpen
  \bibfield  {author} {\bibinfo {author} {\bibfnamefont {P.}~\bibnamefont {Teixeira}}, \bibinfo {author} {\bibfnamefont {A.}~\bibnamefont {Blanco}}, \bibinfo {author} {\bibfnamefont {B.}~\bibnamefont {Caldeira}},  \emph {et~al.},\ }\bibfield  {title} {\enquote {\bibinfo {title} {Muography {Applied} in {Underground} {Geological} {Surveys}: {Ongoing} {Work} at the {Lousal} {Mine} ({Iberian} {Pyrite} {Belt}, {Portugal})},}\ }in\ \href {\doibase 10.1007/978-3-031-48715-6_38} {\emph {\bibinfo {booktitle} {Recent {Research} on {Geotechnical} {Engineering}, {Remote} {Sensing}, {Geophysics} and {Earthquake} {Seismology}}}},\ \bibinfo {editor} {edited by\ \bibinfo {editor} {\bibfnamefont {M.}~\bibnamefont {Bezzeghoud}}, \bibinfo {editor} {\bibfnamefont {Z.~A.}\ \bibnamefont {Ergüler}}, \bibinfo {editor} {\bibfnamefont {J.}~\bibnamefont {Rodrigo-Comino}}, \bibinfo {editor} {\bibfnamefont {M.~K.}\ \bibnamefont {Jat}}, \bibinfo {editor} {\bibfnamefont {R.}~\bibnamefont {Kalatehjari}}, \bibinfo {editor} {\bibfnamefont
  {D.~S.}\ \bibnamefont {Bisht}}, \bibinfo {editor} {\bibfnamefont {A.}~\bibnamefont {Biswas}}, \bibinfo {editor} {\bibfnamefont {H.~I.}\ \bibnamefont {Chaminé}}, \bibinfo {editor} {\bibfnamefont {A.~A.}\ \bibnamefont {Shah}}, \bibinfo {editor} {\bibfnamefont {A.~E.}\ \bibnamefont {Radwan}}, \bibinfo {editor} {\bibfnamefont {J.}~\bibnamefont {Knight}}, \bibinfo {editor} {\bibfnamefont {D.}~\bibnamefont {Panagoulia}}, \bibinfo {editor} {\bibfnamefont {A.}~\bibnamefont {Kallel}}, \bibinfo {editor} {\bibfnamefont {V.}~\bibnamefont {Turan}}, \bibinfo {editor} {\bibfnamefont {H.}~\bibnamefont {Chenchouni}}, \bibinfo {editor} {\bibfnamefont {A.}~\bibnamefont {Ciner}}, \ and\ \bibinfo {editor} {\bibfnamefont {M.}~\bibnamefont {Gentilucci}}}\ (\bibinfo  {publisher} {Springer Nature Switzerland},\ \bibinfo {address} {Cham},\ \bibinfo {year} {2024})\ pp.\ \bibinfo {pages} {173--177}\BibitemShut {NoStop}%
\bibitem [{\citenamefont {Beni}\ \emph {et~al.}(2023)\citenamefont {Beni}, \citenamefont {Borselli}, \citenamefont {Bonechi}, \citenamefont {Bongi} \emph {et~al.}}]{Beni2023}%
  \BibitemOpen
  \bibfield  {author} {\bibinfo {author} {\bibfnamefont {T.}~\bibnamefont {Beni}}, \bibinfo {author} {\bibfnamefont {D.}~\bibnamefont {Borselli}}, \bibinfo {author} {\bibfnamefont {L.}~\bibnamefont {Bonechi}}, \bibinfo {author} {\bibfnamefont {M.}~\bibnamefont {Bongi}},  \emph {et~al.},\ }\bibfield  {title} {\enquote {\bibinfo {title} {Transmission-{Based} {Muography} for {Ore} {Bodies} {Prospecting}: {A} {Case} {Study} from a {Skarn} {Complex} in {Italy}},}\ }\href {\doibase 10.1007/s11053-023-10201-8} {\bibfield  {journal} {\bibinfo  {journal} {Natural Resources Research}\ }\textbf {\bibinfo {volume} {32}},\ \bibinfo {pages} {1529--1547} (\bibinfo {year} {2023})}\BibitemShut {NoStop}%
\bibitem [{\citenamefont {Nishiyama}\ \emph {et~al.}(2014)\citenamefont {Nishiyama}, \citenamefont {Tanaka}, \citenamefont {Okubo}, \citenamefont {Oshima}, \citenamefont {Tanaka},\ and\ \citenamefont {Maekawa}}]{Nishiyama2014}%
  \BibitemOpen
  \bibfield  {author} {\bibinfo {author} {\bibfnamefont {R.}~\bibnamefont {Nishiyama}}, \bibinfo {author} {\bibfnamefont {Y.}~\bibnamefont {Tanaka}}, \bibinfo {author} {\bibfnamefont {S.}~\bibnamefont {Okubo}}, \bibinfo {author} {\bibfnamefont {H.}~\bibnamefont {Oshima}}, \bibinfo {author} {\bibfnamefont {H.~K.~M.}\ \bibnamefont {Tanaka}}, \ and\ \bibinfo {author} {\bibfnamefont {T.}~\bibnamefont {Maekawa}},\ }\bibfield  {title} {\enquote {\bibinfo {title} {Integrated processing of muon radiography and gravity anomaly data toward the realization of high‐resolution 3‐d density structural analysis of volcanoes: Case study of showa‐shinzan lava dome, usu, japan},}\ }\href {\doibase 10.1002/2013jb010234} {\bibfield  {journal} {\bibinfo  {journal} {Journal of Geophysical Research: Solid Earth}\ }\textbf {\bibinfo {volume} {119}},\ \bibinfo {pages} {699–710} (\bibinfo {year} {2014})}\BibitemShut {NoStop}%
\bibitem [{\citenamefont {Nagamine}\ \emph {et~al.}(1995)\citenamefont {Nagamine}, \citenamefont {Iwasaki}, \citenamefont {Shimomura},\ and\ \citenamefont {Ishida}}]{Nagamine:1995np}%
  \BibitemOpen
  \bibfield  {author} {\bibinfo {author} {\bibfnamefont {K.}~\bibnamefont {Nagamine}}, \bibinfo {author} {\bibfnamefont {M.}~\bibnamefont {Iwasaki}}, \bibinfo {author} {\bibfnamefont {K.}~\bibnamefont {Shimomura}}, \ and\ \bibinfo {author} {\bibfnamefont {K.}~\bibnamefont {Ishida}},\ }\bibfield  {title} {\enquote {\bibinfo {title} {{Method of probing inner structure of geophysical substance with the horizontal cosmic ray muons and possible application to volcanic eruption prediction}},}\ }\href {\doibase 10.1016/0168-9002(94)01169-9} {\bibfield  {journal} {\bibinfo  {journal} {Nucl. Instrum. Meth. A}\ }\textbf {\bibinfo {volume} {356}},\ \bibinfo {pages} {585--595} (\bibinfo {year} {1995})}\BibitemShut {NoStop}%
\bibitem [{\citenamefont {Tanaka}\ \emph {et~al.}(2009)\citenamefont {Tanaka}, \citenamefont {Uchida}, \citenamefont {Tanaka}, \citenamefont {Takeo}, \citenamefont {Oikawa}, \citenamefont {Ohminato}, \citenamefont {Aoki}, \citenamefont {Koyama},\ and\ \citenamefont {Tsuji}}]{Tanaka2009}%
  \BibitemOpen
  \bibfield  {author} {\bibinfo {author} {\bibfnamefont {H.~K.~M.}\ \bibnamefont {Tanaka}}, \bibinfo {author} {\bibfnamefont {T.}~\bibnamefont {Uchida}}, \bibinfo {author} {\bibfnamefont {M.}~\bibnamefont {Tanaka}}, \bibinfo {author} {\bibfnamefont {M.}~\bibnamefont {Takeo}}, \bibinfo {author} {\bibfnamefont {J.}~\bibnamefont {Oikawa}}, \bibinfo {author} {\bibfnamefont {T.}~\bibnamefont {Ohminato}}, \bibinfo {author} {\bibfnamefont {Y.}~\bibnamefont {Aoki}}, \bibinfo {author} {\bibfnamefont {E.}~\bibnamefont {Koyama}}, \ and\ \bibinfo {author} {\bibfnamefont {H.}~\bibnamefont {Tsuji}},\ }\bibfield  {title} {\enquote {\bibinfo {title} {Detecting a mass change inside a volcano by cosmic‐ray muon radiography (muography): First results from measurements at asama volcano, japan},}\ }\href {\doibase 10.1029/2009gl039448} {\bibfield  {journal} {\bibinfo  {journal} {Geophysical Research Letters}\ }\textbf {\bibinfo {volume} {36}} (\bibinfo {year} {2009}),\ 10.1029/2009gl039448}\BibitemShut {NoStop}%
\bibitem [{\citenamefont {Alvarez}\ \emph {et~al.}(1970)\citenamefont {Alvarez}, \citenamefont {Anderson}, \citenamefont {Bedwei}, \citenamefont {Burkhard}, \citenamefont {Fakhry}, \citenamefont {Girgis}, \citenamefont {Goneid}, \citenamefont {Hassan}, \citenamefont {Iverson}, \citenamefont {Lynch}, \citenamefont {Miligy}, \citenamefont {Moussa}, \citenamefont {Sharkawi},\ and\ \citenamefont {Yazolino}}]{Alvarez1970}%
  \BibitemOpen
  \bibfield  {author} {\bibinfo {author} {\bibfnamefont {L.~W.}\ \bibnamefont {Alvarez}}, \bibinfo {author} {\bibfnamefont {J.~A.}\ \bibnamefont {Anderson}}, \bibinfo {author} {\bibfnamefont {F.~E.}\ \bibnamefont {Bedwei}}, \bibinfo {author} {\bibfnamefont {J.}~\bibnamefont {Burkhard}}, \bibinfo {author} {\bibfnamefont {A.}~\bibnamefont {Fakhry}}, \bibinfo {author} {\bibfnamefont {A.}~\bibnamefont {Girgis}}, \bibinfo {author} {\bibfnamefont {A.}~\bibnamefont {Goneid}}, \bibinfo {author} {\bibfnamefont {F.}~\bibnamefont {Hassan}}, \bibinfo {author} {\bibfnamefont {D.}~\bibnamefont {Iverson}}, \bibinfo {author} {\bibfnamefont {G.}~\bibnamefont {Lynch}}, \bibinfo {author} {\bibfnamefont {Z.}~\bibnamefont {Miligy}}, \bibinfo {author} {\bibfnamefont {A.~H.}\ \bibnamefont {Moussa}}, \bibinfo {author} {\bibfnamefont {M.}~\bibnamefont {Sharkawi}}, \ and\ \bibinfo {author} {\bibfnamefont {L.}~\bibnamefont {Yazolino}},\ }\bibfield  {title} {\enquote {\bibinfo {title} {Search for hidden chambers in the pyramids: The
  structure of the second pyramid of giza is determined by cosmic-ray absorption.}}\ }\href {\doibase 10.1126/science.167.3919.832} {\bibfield  {journal} {\bibinfo  {journal} {Science}\ }\textbf {\bibinfo {volume} {167}},\ \bibinfo {pages} {832–839} (\bibinfo {year} {1970})}\BibitemShut {NoStop}%
\bibitem [{\citenamefont {Liu}\ \emph {et~al.}(2023)\citenamefont {Liu}, \citenamefont {Luo}, \citenamefont {Tian}, \citenamefont {Yao}, \citenamefont {Niu}, \citenamefont {Jin}, \citenamefont {Gao}, \citenamefont {Rong}, \citenamefont {Fu}, \citenamefont {Kang}, \citenamefont {Fu}, \citenamefont {Wu}, \citenamefont {Gao}, \citenamefont {Gong}, \citenamefont {Zhang}, \citenamefont {Luo}, \citenamefont {Liu}, \citenamefont {Tian}, \citenamefont {Yu}, \citenamefont {Wu}, \citenamefont {Chen}, \citenamefont {Liu},\ and\ \citenamefont {Liu}}]{Liu2023}%
  \BibitemOpen
  \bibfield  {author} {\bibinfo {author} {\bibfnamefont {G.}~\bibnamefont {Liu}}, \bibinfo {author} {\bibfnamefont {X.}~\bibnamefont {Luo}}, \bibinfo {author} {\bibfnamefont {H.}~\bibnamefont {Tian}}, \bibinfo {author} {\bibfnamefont {K.}~\bibnamefont {Yao}}, \bibinfo {author} {\bibfnamefont {F.}~\bibnamefont {Niu}}, \bibinfo {author} {\bibfnamefont {L.}~\bibnamefont {Jin}}, \bibinfo {author} {\bibfnamefont {J.}~\bibnamefont {Gao}}, \bibinfo {author} {\bibfnamefont {J.}~\bibnamefont {Rong}}, \bibinfo {author} {\bibfnamefont {Z.}~\bibnamefont {Fu}}, \bibinfo {author} {\bibfnamefont {Y.}~\bibnamefont {Kang}}, \bibinfo {author} {\bibfnamefont {Y.}~\bibnamefont {Fu}}, \bibinfo {author} {\bibfnamefont {C.}~\bibnamefont {Wu}}, \bibinfo {author} {\bibfnamefont {H.}~\bibnamefont {Gao}}, \bibinfo {author} {\bibfnamefont {J.}~\bibnamefont {Gong}}, \bibinfo {author} {\bibfnamefont {W.}~\bibnamefont {Zhang}}, \bibinfo {author} {\bibfnamefont {X.}~\bibnamefont {Luo}}, \bibinfo {author} {\bibfnamefont {C.}~\bibnamefont
  {Liu}}, \bibinfo {author} {\bibfnamefont {X.}~\bibnamefont {Tian}}, \bibinfo {author} {\bibfnamefont {M.}~\bibnamefont {Yu}}, \bibinfo {author} {\bibfnamefont {F.}~\bibnamefont {Wu}}, \bibinfo {author} {\bibfnamefont {J.}~\bibnamefont {Chen}}, \bibinfo {author} {\bibfnamefont {J.}~\bibnamefont {Liu}}, \ and\ \bibinfo {author} {\bibfnamefont {Z.}~\bibnamefont {Liu}},\ }\bibfield  {title} {\enquote {\bibinfo {title} {High-precision muography in archaeogeophysics: A case study on xi’an defensive walls},}\ }\href {\doibase 10.1063/5.0123337} {\bibfield  {journal} {\bibinfo  {journal} {Journal of Applied Physics}\ }\textbf {\bibinfo {volume} {133}} (\bibinfo {year} {2023}),\ 10.1063/5.0123337}\BibitemShut {NoStop}%
\bibitem [{\citenamefont {Morishima}\ \emph {et~al.}(2017)\citenamefont {Morishima}, \citenamefont {Kuno}, \citenamefont {Nishio} \emph {et~al.}}]{Morishima2017}%
  \BibitemOpen
  \bibfield  {author} {\bibinfo {author} {\bibfnamefont {K.}~\bibnamefont {Morishima}}, \bibinfo {author} {\bibfnamefont {M.}~\bibnamefont {Kuno}}, \bibinfo {author} {\bibfnamefont {A.}~\bibnamefont {Nishio}},  \emph {et~al.},\ }\bibfield  {title} {\enquote {\bibinfo {title} {Discovery of a big void in khufu’s pyramid by observation of cosmic-ray muons},}\ }\href@noop {} {\bibfield  {journal} {\bibinfo  {journal} {Nature}\ } (\bibinfo {year} {2017})}\BibitemShut {NoStop}%
\bibitem [{\citenamefont {Tanaka}\ \emph {et~al.}(2022{\natexlab{a}})\citenamefont {Tanaka}, \citenamefont {Gluyas}, \citenamefont {Holma}, \citenamefont {Joutsenvaara}, \citenamefont {Kuusiniemi}, \citenamefont {Leone}, \citenamefont {Lo~Presti}, \citenamefont {Matsushima}, \citenamefont {Ol{\'a}h}, \citenamefont {Steigerwald} \emph {et~al.}}]{Tanaka2022a}%
  \BibitemOpen
  \bibfield  {author} {\bibinfo {author} {\bibfnamefont {H.~K.}\ \bibnamefont {Tanaka}}, \bibinfo {author} {\bibfnamefont {J.}~\bibnamefont {Gluyas}}, \bibinfo {author} {\bibfnamefont {M.}~\bibnamefont {Holma}}, \bibinfo {author} {\bibfnamefont {J.}~\bibnamefont {Joutsenvaara}}, \bibinfo {author} {\bibfnamefont {P.}~\bibnamefont {Kuusiniemi}}, \bibinfo {author} {\bibfnamefont {G.}~\bibnamefont {Leone}}, \bibinfo {author} {\bibfnamefont {D.}~\bibnamefont {Lo~Presti}}, \bibinfo {author} {\bibfnamefont {J.}~\bibnamefont {Matsushima}}, \bibinfo {author} {\bibfnamefont {L.}~\bibnamefont {Ol{\'a}h}}, \bibinfo {author} {\bibfnamefont {S.}~\bibnamefont {Steigerwald}},  \emph {et~al.},\ }\bibfield  {title} {\enquote {\bibinfo {title} {Atmospheric muography for imaging and monitoring tropic cyclones},}\ }\href@noop {} {\bibfield  {journal} {\bibinfo  {journal} {Scientific reports}\ }\textbf {\bibinfo {volume} {12}},\ \bibinfo {pages} {1--14} (\bibinfo {year} {2022}{\natexlab{a}})}\BibitemShut {NoStop}%
\bibitem [{\citenamefont {Tanaka}\ \emph {et~al.}(2021)\citenamefont {Tanaka}, \citenamefont {Aichi}, \citenamefont {Bozza}, \citenamefont {Coniglione} \emph {et~al.}}]{Tanaka2021a}%
  \BibitemOpen
  \bibfield  {author} {\bibinfo {author} {\bibfnamefont {H.}~\bibnamefont {Tanaka}}, \bibinfo {author} {\bibfnamefont {M.}~\bibnamefont {Aichi}}, \bibinfo {author} {\bibfnamefont {C.}~\bibnamefont {Bozza}}, \bibinfo {author} {\bibfnamefont {R.}~\bibnamefont {Coniglione}},  \emph {et~al.},\ }\bibfield  {title} {\enquote {\bibinfo {title} {First results of undersea muography with the tokyo-bay seafloor hyper-kilometric submarine deep detector},}\ }\href {\doibase 10.1038/s41598-021-98559-8} {\bibfield  {journal} {\bibinfo  {journal} {Scientific Reports}\ }\textbf {\bibinfo {volume} {11}} (\bibinfo {year} {2021}),\ 10.1038/s41598-021-98559-8}\BibitemShut {NoStop}%
\bibitem [{\citenamefont {Tanaka}\ \emph {et~al.}(2022{\natexlab{b}})\citenamefont {Tanaka}, \citenamefont {Aichi}, \citenamefont {Balogh}, \citenamefont {Bozza}, \citenamefont {Coniglione}, \citenamefont {Gluyas}, \citenamefont {Hayashi}, \citenamefont {Holma}, \citenamefont {Joutsenvaara}, \citenamefont {Kamoshida}, \citenamefont {Kato}, \citenamefont {Kin}, \citenamefont {Kuusiniemi}, \citenamefont {Leone}, \citenamefont {Lo~Presti}, \citenamefont {Matsushima}, \citenamefont {Miyamoto}, \citenamefont {Mori}, \citenamefont {Nomura},\ and\ \citenamefont {Varga}}]{Tanaka2022}%
  \BibitemOpen
  \bibfield  {author} {\bibinfo {author} {\bibfnamefont {H.}~\bibnamefont {Tanaka}}, \bibinfo {author} {\bibfnamefont {M.}~\bibnamefont {Aichi}}, \bibinfo {author} {\bibfnamefont {S.}~\bibnamefont {Balogh}}, \bibinfo {author} {\bibfnamefont {C.}~\bibnamefont {Bozza}}, \bibinfo {author} {\bibfnamefont {R.}~\bibnamefont {Coniglione}}, \bibinfo {author} {\bibfnamefont {J.}~\bibnamefont {Gluyas}}, \bibinfo {author} {\bibfnamefont {N.}~\bibnamefont {Hayashi}}, \bibinfo {author} {\bibfnamefont {M.}~\bibnamefont {Holma}}, \bibinfo {author} {\bibfnamefont {J.}~\bibnamefont {Joutsenvaara}}, \bibinfo {author} {\bibfnamefont {O.}~\bibnamefont {Kamoshida}}, \bibinfo {author} {\bibfnamefont {Y.}~\bibnamefont {Kato}}, \bibinfo {author} {\bibfnamefont {T.}~\bibnamefont {Kin}}, \bibinfo {author} {\bibfnamefont {P.}~\bibnamefont {Kuusiniemi}}, \bibinfo {author} {\bibfnamefont {G.}~\bibnamefont {Leone}}, \bibinfo {author} {\bibfnamefont {D.}~\bibnamefont {Lo~Presti}}, \bibinfo {author} {\bibfnamefont {J.}~\bibnamefont
  {Matsushima}}, \bibinfo {author} {\bibfnamefont {H.}~\bibnamefont {Miyamoto}}, \bibinfo {author} {\bibfnamefont {H.}~\bibnamefont {Mori}}, \bibinfo {author} {\bibfnamefont {Y.}~\bibnamefont {Nomura}}, \ and\ \bibinfo {author} {\bibfnamefont {D.}~\bibnamefont {Varga}},\ }\bibfield  {title} {\enquote {\bibinfo {title} {Periodic sea-level oscillation in {Tokyo} {Bay} detected with the {Tokyo}-{Bay} seafloor hyper-kilometric submarine deep detector ({TS}-{HKMSDD})},}\ }\href {\doibase 10.1038/s41598-022-10078-2} {\bibfield  {journal} {\bibinfo  {journal} {Scientific Reports}\ }\textbf {\bibinfo {volume} {12}},\ \bibinfo {pages} {6097} (\bibinfo {year} {2022}{\natexlab{b}})}\BibitemShut {NoStop}%
\bibitem [{\citenamefont {Tanaka}(2020)}]{Tanaka2020}%
  \BibitemOpen
  \bibfield  {author} {\bibinfo {author} {\bibfnamefont {H.}~\bibnamefont {Tanaka}},\ }\bibfield  {title} {\enquote {\bibinfo {title} {Muometric positioning system (mups) with cosmic muons as a new underwater and underground positioning technique},}\ }\href {\doibase 10.1038/s41598-020-75843-7} {\bibfield  {journal} {\bibinfo  {journal} {Scientific Reports}\ }\textbf {\bibinfo {volume} {10}},\ \bibinfo {pages} {18896} (\bibinfo {year} {2020})}\BibitemShut {NoStop}%
\bibitem [{\citenamefont {Varga}\ and\ \citenamefont {Tanaka}(2024)}]{Varga2024a}%
  \BibitemOpen
  \bibfield  {author} {\bibinfo {author} {\bibfnamefont {D.}~\bibnamefont {Varga}}\ and\ \bibinfo {author} {\bibfnamefont {H.~K.~M.}\ \bibnamefont {Tanaka}},\ }\bibfield  {title} {\enquote {\bibinfo {title} {Developments of a centimeter-level precise muometric wireless navigation system ({MuWNS}-{V}) and its first demonstration using directional information from tracking detectors},}\ }\href {\doibase 10.1038/s41598-024-57857-7} {\bibfield  {journal} {\bibinfo  {journal} {Scientific Reports}\ }\textbf {\bibinfo {volume} {14}},\ \bibinfo {pages} {7605} (\bibinfo {year} {2024})}\BibitemShut {NoStop}%
\bibitem [{\citenamefont {Chilingarian}, \citenamefont {Chilingaryan},\ and\ \citenamefont {Zazyan}(2024)}]{Chilingarian2024}%
  \BibitemOpen
  \bibfield  {author} {\bibinfo {author} {\bibfnamefont {A.}~\bibnamefont {Chilingarian}}, \bibinfo {author} {\bibfnamefont {S.}~\bibnamefont {Chilingaryan}}, \ and\ \bibinfo {author} {\bibfnamefont {M.}~\bibnamefont {Zazyan}},\ }\href {\doibase 10.48550/arXiv.2406.18608} {\enquote {\bibinfo {title} {Cosmic {Ray} {Navigation} {System} ({CRoNS}) for {Autonomous} {Navigation} in {GPS}-{Denied} {Environments}},}\ }\bibinfo {type} {Tech. Rep.}\ (\bibinfo {year} {2024})\ \bibinfo {note} {arXiv:2406.18608 [physics] type: article}\BibitemShut {NoStop}%
\bibitem [{\citenamefont {Thompson}\ \emph {et~al.}(2020)\citenamefont {Thompson}, \citenamefont {Stowell}, \citenamefont {Fargher}, \citenamefont {Steer}, \citenamefont {Loughney}, \citenamefont {O'Sullivan}, \citenamefont {Gluyas}, \citenamefont {Blaney},\ and\ \citenamefont {Pidcock}}]{Thompson2020}%
  \BibitemOpen
  \bibfield  {author} {\bibinfo {author} {\bibfnamefont {L.~F.}\ \bibnamefont {Thompson}}, \bibinfo {author} {\bibfnamefont {J.~P.}\ \bibnamefont {Stowell}}, \bibinfo {author} {\bibfnamefont {S.~J.}\ \bibnamefont {Fargher}}, \bibinfo {author} {\bibfnamefont {C.~A.}\ \bibnamefont {Steer}}, \bibinfo {author} {\bibfnamefont {K.~L.}\ \bibnamefont {Loughney}}, \bibinfo {author} {\bibfnamefont {E.~M.}\ \bibnamefont {O'Sullivan}}, \bibinfo {author} {\bibfnamefont {J.~G.}\ \bibnamefont {Gluyas}}, \bibinfo {author} {\bibfnamefont {S.~W.}\ \bibnamefont {Blaney}}, \ and\ \bibinfo {author} {\bibfnamefont {R.~J.}\ \bibnamefont {Pidcock}},\ }\bibfield  {title} {\enquote {\bibinfo {title} {Muon tomography for railway tunnel imaging},}\ }\href {\doibase 10.1103/PhysRevResearch.2.023017} {\bibfield  {journal} {\bibinfo  {journal} {Phys. Rev. Res.}\ }\textbf {\bibinfo {volume} {2}},\ \bibinfo {pages} {023017} (\bibinfo {year} {2020})}\BibitemShut {NoStop}%
\bibitem [{\citenamefont {Cohu}\ \emph {et~al.}(2023)\citenamefont {Cohu}, \citenamefont {Chevalier}, \citenamefont {Nechyporuk}, \citenamefont {Franzen}, \citenamefont {Sauerwald}, \citenamefont {Ianigro},\ and\ \citenamefont {Marteau}}]{Cohu2023}%
  \BibitemOpen
  \bibfield  {author} {\bibinfo {author} {\bibfnamefont {A.}~\bibnamefont {Cohu}}, \bibinfo {author} {\bibfnamefont {A.}~\bibnamefont {Chevalier}}, \bibinfo {author} {\bibfnamefont {O.}~\bibnamefont {Nechyporuk}}, \bibinfo {author} {\bibfnamefont {A.}~\bibnamefont {Franzen}}, \bibinfo {author} {\bibfnamefont {J.}~\bibnamefont {Sauerwald}}, \bibinfo {author} {\bibfnamefont {J.-C.}\ \bibnamefont {Ianigro}}, \ and\ \bibinfo {author} {\bibfnamefont {J.}~\bibnamefont {Marteau}},\ }\bibfield  {title} {\enquote {\bibinfo {title} {First {3D} reconstruction of a blast furnace using muography},}\ }\href {\doibase 10.1088/1748-0221/18/07/P07004} {\bibfield  {journal} {\bibinfo  {journal} {Journal of Instrumentation}\ }\textbf {\bibinfo {volume} {18}},\ \bibinfo {pages} {P07004} (\bibinfo {year} {2023})},\ \bibinfo {note} {arXiv:2301.04354 [hep-ex, physics:hep-ph, physics:physics]}\BibitemShut {NoStop}%
\bibitem [{\citenamefont {Borozdin}\ \emph {et~al.}(2012)\citenamefont {Borozdin}, \citenamefont {Greene}, \citenamefont {Luki{\'c}}, \citenamefont {Milner}, \citenamefont {Miyadera}, \citenamefont {Morris},\ and\ \citenamefont {Perry}}]{Borozdin2012}%
  \BibitemOpen
  \bibfield  {author} {\bibinfo {author} {\bibfnamefont {K.~N.}\ \bibnamefont {Borozdin}}, \bibinfo {author} {\bibfnamefont {S.~J.}\ \bibnamefont {Greene}}, \bibinfo {author} {\bibfnamefont {Z.}~\bibnamefont {Luki{\'c}}}, \bibinfo {author} {\bibfnamefont {E.~C.}\ \bibnamefont {Milner}}, \bibinfo {author} {\bibfnamefont {H.}~\bibnamefont {Miyadera}}, \bibinfo {author} {\bibfnamefont {C.}~\bibnamefont {Morris}}, \ and\ \bibinfo {author} {\bibfnamefont {J.}~\bibnamefont {Perry}},\ }\bibfield  {title} {\enquote {\bibinfo {title} {Cosmic ray radiography of the damaged cores of the fukushima reactors.}}\ }\href@noop {} {\bibfield  {journal} {\bibinfo  {journal} {Physical Review Letters}\ } (\bibinfo {year} {2012})}\BibitemShut {NoStop}%
\bibitem [{\citenamefont {Procureur}\ \emph {et~al.}(2023)\citenamefont {Procureur}, \citenamefont {Attié}, \citenamefont {Gallego}, \citenamefont {Gomez}, \citenamefont {Gonzales}, \citenamefont {Lefèvre}, \citenamefont {Lehuraux}, \citenamefont {Lesage}, \citenamefont {Mandjavidze}, \citenamefont {Mas},\ and\ \citenamefont {Pomarède}}]{Procureur2023}%
  \BibitemOpen
  \bibfield  {author} {\bibinfo {author} {\bibfnamefont {S.}~\bibnamefont {Procureur}}, \bibinfo {author} {\bibfnamefont {D.}~\bibnamefont {Attié}}, \bibinfo {author} {\bibfnamefont {L.}~\bibnamefont {Gallego}}, \bibinfo {author} {\bibfnamefont {H.}~\bibnamefont {Gomez}}, \bibinfo {author} {\bibfnamefont {P.}~\bibnamefont {Gonzales}}, \bibinfo {author} {\bibfnamefont {B.}~\bibnamefont {Lefèvre}}, \bibinfo {author} {\bibfnamefont {M.}~\bibnamefont {Lehuraux}}, \bibinfo {author} {\bibfnamefont {B.}~\bibnamefont {Lesage}}, \bibinfo {author} {\bibfnamefont {I.}~\bibnamefont {Mandjavidze}}, \bibinfo {author} {\bibfnamefont {P.}~\bibnamefont {Mas}}, \ and\ \bibinfo {author} {\bibfnamefont {D.}~\bibnamefont {Pomarède}},\ }\bibfield  {title} {\enquote {\bibinfo {title} {{3D} imaging of a nuclear reactor using muography measurements},}\ }\href {\doibase 10.1126/sciadv.abq8431} {\bibfield  {journal} {\bibinfo  {journal} {Science Advances}\ }\textbf {\bibinfo {volume} {9}},\ \bibinfo {pages} {eabq8431} (\bibinfo
  {year} {2023})}\BibitemShut {NoStop}%
\bibitem [{\citenamefont {Morris}\ \emph {et~al.}(2025)\citenamefont {Morris}, \citenamefont {Perry}, \citenamefont {Borozdin}, \citenamefont {Bacon}, \citenamefont {Durham}, \citenamefont {Freeman}, \citenamefont {Merrill},\ and\ \citenamefont {Miyadera}}]{Morris2025}%
  \BibitemOpen
  \bibfield  {author} {\bibinfo {author} {\bibfnamefont {C.}~\bibnamefont {Morris}}, \bibinfo {author} {\bibfnamefont {J.}~\bibnamefont {Perry}}, \bibinfo {author} {\bibfnamefont {K.}~\bibnamefont {Borozdin}}, \bibinfo {author} {\bibfnamefont {J.}~\bibnamefont {Bacon}}, \bibinfo {author} {\bibfnamefont {J.~M.}\ \bibnamefont {Durham}}, \bibinfo {author} {\bibfnamefont {M.~S.}\ \bibnamefont {Freeman}}, \bibinfo {author} {\bibfnamefont {F.~E.}\ \bibnamefont {Merrill}}, \ and\ \bibinfo {author} {\bibfnamefont {H.}~\bibnamefont {Miyadera}},\ }\bibfield  {title} {\enquote {\bibinfo {title} {Cosmic ray radiography of a human phantom},}\ }\href {\doibase 10.1063/5.0253548} {\bibfield  {journal} {\bibinfo  {journal} {Journal of Applied Physics}\ }\textbf {\bibinfo {volume} {137}} (\bibinfo {year} {2025}),\ 10.1063/5.0253548}\BibitemShut {NoStop}%
\bibitem [{\citenamefont {Nakamura}\ \emph {et~al.}(2006)\citenamefont {Nakamura}, \citenamefont {Ariga}, \citenamefont {Ban}, \citenamefont {Fukuda} \emph {et~al.}}]{Nakamura2006}%
  \BibitemOpen
  \bibfield  {author} {\bibinfo {author} {\bibfnamefont {T.}~\bibnamefont {Nakamura}}, \bibinfo {author} {\bibfnamefont {A.}~\bibnamefont {Ariga}}, \bibinfo {author} {\bibfnamefont {T.}~\bibnamefont {Ban}}, \bibinfo {author} {\bibfnamefont {T.}~\bibnamefont {Fukuda}},  \emph {et~al.},\ }\bibfield  {title} {\enquote {\bibinfo {title} {The {OPERA} film: {New} nuclear emulsion for large-scale, high-precision experiments},}\ }\href {\doibase 10.1016/j.nima.2005.08.109} {\bibfield  {journal} {\bibinfo  {journal} {Nuclear Instruments and Methods in Physics Research Section A: Accelerators, Spectrometers, Detectors and Associated Equipment}\ }\textbf {\bibinfo {volume} {556}},\ \bibinfo {pages} {80--86} (\bibinfo {year} {2006})}\BibitemShut {NoStop}%
\bibitem [{\citenamefont {Gnanvo}\ \emph {et~al.}(2011)\citenamefont {Gnanvo}, \citenamefont {Grasso}, \citenamefont {Hohlmann}, \citenamefont {Locke}, \citenamefont {Quintero},\ and\ \citenamefont {Mitra}}]{Gnanvo2011}%
  \BibitemOpen
  \bibfield  {author} {\bibinfo {author} {\bibfnamefont {K.}~\bibnamefont {Gnanvo}}, \bibinfo {author} {\bibfnamefont {L.~V.}\ \bibnamefont {Grasso}}, \bibinfo {author} {\bibfnamefont {M.}~\bibnamefont {Hohlmann}}, \bibinfo {author} {\bibfnamefont {J.~B.}\ \bibnamefont {Locke}}, \bibinfo {author} {\bibfnamefont {A.}~\bibnamefont {Quintero}}, \ and\ \bibinfo {author} {\bibfnamefont {D.}~\bibnamefont {Mitra}},\ }\bibfield  {title} {\enquote {\bibinfo {title} {Imaging of high-\textit{{Z}} material for nuclear contraband detection with a minimal prototype of a muon tomography station based on {GEM} detectors},}\ }\href {\doibase 10.1016/j.nima.2011.01.163} {\bibfield  {journal} {\bibinfo  {journal} {Nuclear Instruments and Methods in Physics Research Section A: Accelerators, Spectrometers, Detectors and Associated Equipment}\ }\bibinfo {series} {Symposium on {Radiation} {Measurements} and {Applications} ({SORMA}) {XII} 2010},\ \textbf {\bibinfo {volume} {652}},\ \bibinfo {pages} {16--20} (\bibinfo {year}
  {2011})}\BibitemShut {NoStop}%
\bibitem [{jan(2024)}]{jan2024}%
  \BibitemOpen
  \href {\doibase 10.48550/arXiv.2401.07247} {\enquote {\bibinfo {title} {Exploring {Advanced} {Detector} {Technologies} for {Muon} {Radiography} {Applications}},}\ } (\bibinfo {year} {2024}),\ \bibinfo {note} {arXiv:2401.07247 [physics]}\BibitemShut {NoStop}%
\bibitem [{\citenamefont {Niculescu-Oglinzanu}\ \emph {et~al.}(2024)\citenamefont {Niculescu-Oglinzanu}, \citenamefont {Stanca}, \citenamefont {Bălăceanu}, \citenamefont {Dobre}, \citenamefont {Gherghel-Lascu}, \citenamefont {Saftoiu}, \citenamefont {Smău},\ and\ \citenamefont {Vancea}}]{NiculescuOglinzanu2024}%
  \BibitemOpen
  \bibfield  {author} {\bibinfo {author} {\bibfnamefont {M.}~\bibnamefont {Niculescu-Oglinzanu}}, \bibinfo {author} {\bibfnamefont {D.}~\bibnamefont {Stanca}}, \bibinfo {author} {\bibfnamefont {A.}~\bibnamefont {Bălăceanu}}, \bibinfo {author} {\bibfnamefont {M.}~\bibnamefont {Dobre}}, \bibinfo {author} {\bibfnamefont {A.}~\bibnamefont {Gherghel-Lascu}}, \bibinfo {author} {\bibfnamefont {A.}~\bibnamefont {Saftoiu}}, \bibinfo {author} {\bibfnamefont {R.}~\bibnamefont {Smău}}, \ and\ \bibinfo {author} {\bibfnamefont {C.}~\bibnamefont {Vancea}},\ }\bibfield  {title} {\enquote {\bibinfo {title} {{SiRO}, a scintillator-based hodoscope for muography applications},}\ }\href {\doibase 10.1063/5.0224843} {\bibfield  {journal} {\bibinfo  {journal} {Journal of Applied Physics}\ }\textbf {\bibinfo {volume} {136}},\ \bibinfo {pages} {174501} (\bibinfo {year} {2024})}\BibitemShut {NoStop}%
\bibitem [{\citenamefont {Ning}\ \emph {et~al.}(2025)\citenamefont {Ning}, \citenamefont {Yuan}, \citenamefont {Yu}, \citenamefont {Chen}, \citenamefont {Xie}, \citenamefont {Jiang}, \citenamefont {Liu}, \citenamefont {Lu}, \citenamefont {Sun}, \citenamefont {Chen},\ and\ \citenamefont {Tang}}]{ning2025developmentportablecosmicraymuon}%
  \BibitemOpen
  \bibfield  {author} {\bibinfo {author} {\bibfnamefont {Y.}~\bibnamefont {Ning}}, \bibinfo {author} {\bibfnamefont {Y.}~\bibnamefont {Yuan}}, \bibinfo {author} {\bibfnamefont {T.}~\bibnamefont {Yu}}, \bibinfo {author} {\bibfnamefont {H.}~\bibnamefont {Chen}}, \bibinfo {author} {\bibfnamefont {C.}~\bibnamefont {Xie}}, \bibinfo {author} {\bibfnamefont {H.}~\bibnamefont {Jiang}}, \bibinfo {author} {\bibfnamefont {H.}~\bibnamefont {Liu}}, \bibinfo {author} {\bibfnamefont {G.}~\bibnamefont {Lu}}, \bibinfo {author} {\bibfnamefont {M.}~\bibnamefont {Sun}}, \bibinfo {author} {\bibfnamefont {Y.}~\bibnamefont {Chen}}, \ and\ \bibinfo {author} {\bibfnamefont {J.}~\bibnamefont {Tang}},\ }\href {https://arxiv.org/abs/2503.18800} {\enquote {\bibinfo {title} {Development of portable cosmic-ray muon detector array for muography},}\ } (\bibinfo {year} {2025}),\ \Eprint {http://arxiv.org/abs/2503.18800} {arXiv:2503.18800 [physics.ins-det]} \BibitemShut {NoStop}%
\bibitem [{\citenamefont {Morishima}(2015)}]{Morishima2015}%
  \BibitemOpen
  \bibfield  {author} {\bibinfo {author} {\bibfnamefont {K.}~\bibnamefont {Morishima}},\ }\bibfield  {title} {\enquote {\bibinfo {title} {Latest {Developments} in {Nuclear} {Emulsion} {Technology}},}\ }\href {\doibase 10.1016/j.phpro.2015.11.082} {\bibfield  {journal} {\bibinfo  {journal} {Physics Procedia}\ }\bibinfo {series} {26th {International} {Conference} on {Nuclear} {Tracks} in {Solids} ({ICNTS26}) {Kobe}, {Japan} 15th – 19th {September} 2014},\ \textbf {\bibinfo {volume} {80}},\ \bibinfo {pages} {19--24} (\bibinfo {year} {2015})}\BibitemShut {NoStop}%
\bibitem [{\citenamefont {Nishio}\ \emph {et~al.}(2015)\citenamefont {Nishio}, \citenamefont {Morishima}, \citenamefont {Kuwabara},\ and\ \citenamefont {Nakamura}}]{Nishio2015}%
  \BibitemOpen
  \bibfield  {author} {\bibinfo {author} {\bibfnamefont {A.}~\bibnamefont {Nishio}}, \bibinfo {author} {\bibfnamefont {K.}~\bibnamefont {Morishima}}, \bibinfo {author} {\bibfnamefont {K.}~\bibnamefont {Kuwabara}}, \ and\ \bibinfo {author} {\bibfnamefont {M.}~\bibnamefont {Nakamura}},\ }\bibfield  {title} {\enquote {\bibinfo {title} {Development of {Nuclear} {Emulsion} {Detector} for {Muon} {Radiography}},}\ }\href {\doibase 10.1016/j.phpro.2015.11.084} {\bibfield  {journal} {\bibinfo  {journal} {Physics Procedia}\ }\bibinfo {series} {26th {International} {Conference} on {Nuclear} {Tracks} in {Solids} ({ICNTS26}) {Kobe}, {Japan} 15th – 19th {September} 2014},\ \textbf {\bibinfo {volume} {80}},\ \bibinfo {pages} {74--77} (\bibinfo {year} {2015})}\BibitemShut {NoStop}%
\bibitem [{\citenamefont {Feng}\ \emph {et~al.}(2021)\citenamefont {Feng}, \citenamefont {Zhang}, \citenamefont {Liu}, \citenamefont {Qi}, \citenamefont {Wang}, \citenamefont {Shao},\ and\ \citenamefont {Zhou}}]{Feng2021}%
  \BibitemOpen
  \bibfield  {author} {\bibinfo {author} {\bibfnamefont {J.}~\bibnamefont {Feng}}, \bibinfo {author} {\bibfnamefont {Z.}~\bibnamefont {Zhang}}, \bibinfo {author} {\bibfnamefont {J.}~\bibnamefont {Liu}}, \bibinfo {author} {\bibfnamefont {B.}~\bibnamefont {Qi}}, \bibinfo {author} {\bibfnamefont {A.}~\bibnamefont {Wang}}, \bibinfo {author} {\bibfnamefont {M.}~\bibnamefont {Shao}}, \ and\ \bibinfo {author} {\bibfnamefont {Y.}~\bibnamefont {Zhou}},\ }\bibfield  {title} {\enquote {\bibinfo {title} {A thermal bonding method for manufacturing {Micromegas} detectors},}\ }\href {\doibase 10.1016/j.nima.2020.164958} {\bibfield  {journal} {\bibinfo  {journal} {Nuclear Instruments and Methods in Physics Research Section A: Accelerators, Spectrometers, Detectors and Associated Equipment}\ }\textbf {\bibinfo {volume} {989}},\ \bibinfo {pages} {164958} (\bibinfo {year} {2021})}\BibitemShut {NoStop}%
\bibitem [{\citenamefont {Derré}\ \emph {et~al.}(2001)\citenamefont {Derré}, \citenamefont {Giomataris}, \citenamefont {Zaccone}, \citenamefont {Bay}, \citenamefont {Perroud},\ and\ \citenamefont {Ronga}}]{Derre2001}%
  \BibitemOpen
  \bibfield  {author} {\bibinfo {author} {\bibfnamefont {J.}~\bibnamefont {Derré}}, \bibinfo {author} {\bibfnamefont {Y.}~\bibnamefont {Giomataris}}, \bibinfo {author} {\bibfnamefont {H.}~\bibnamefont {Zaccone}}, \bibinfo {author} {\bibfnamefont {A.}~\bibnamefont {Bay}}, \bibinfo {author} {\bibfnamefont {J.~P.}\ \bibnamefont {Perroud}}, \ and\ \bibinfo {author} {\bibfnamefont {F.}~\bibnamefont {Ronga}},\ }\bibfield  {title} {\enquote {\bibinfo {title} {Spatial resolution in {Micromegas} detectors},}\ }\href {\doibase 10.1016/S0168-9002(00)01051-2} {\bibfield  {journal} {\bibinfo  {journal} {Nuclear Instruments and Methods in Physics Research Section A: Accelerators, Spectrometers, Detectors and Associated Equipment}\ }\textbf {\bibinfo {volume} {459}},\ \bibinfo {pages} {523--531} (\bibinfo {year} {2001})}\BibitemShut {NoStop}%
\bibitem [{\citenamefont {Bugg}, \citenamefont {Efremenko},\ and\ \citenamefont {Vasilyev}(2014)}]{Bugg2014}%
  \BibitemOpen
  \bibfield  {author} {\bibinfo {author} {\bibfnamefont {W.}~\bibnamefont {Bugg}}, \bibinfo {author} {\bibfnamefont {Y.}~\bibnamefont {Efremenko}}, \ and\ \bibinfo {author} {\bibfnamefont {S.}~\bibnamefont {Vasilyev}},\ }\bibfield  {title} {\enquote {\bibinfo {title} {Large plastic scintillator panels with {WLS} fiber readout: {Optimization} of components},}\ }\href {\doibase 10.1016/j.nima.2014.05.055} {\bibfield  {journal} {\bibinfo  {journal} {Nuclear Instruments and Methods in Physics Research Section A: Accelerators, Spectrometers, Detectors and Associated Equipment}\ }\textbf {\bibinfo {volume} {758}},\ \bibinfo {pages} {91--96} (\bibinfo {year} {2014})}\BibitemShut {NoStop}%
\bibitem [{\citenamefont {He}\ \emph {et~al.}(2024)\citenamefont {He}, \citenamefont {Luo}, \citenamefont {Liu}, \citenamefont {Zou}, \citenamefont {Zhang}, \citenamefont {Xiao}, \citenamefont {Huang},\ and\ \citenamefont {Wang}}]{He2024}%
  \BibitemOpen
  \bibfield  {author} {\bibinfo {author} {\bibfnamefont {L.}~\bibnamefont {He}}, \bibinfo {author} {\bibfnamefont {S.-Y.}\ \bibnamefont {Luo}}, \bibinfo {author} {\bibfnamefont {X.-M.}\ \bibnamefont {Liu}}, \bibinfo {author} {\bibfnamefont {Y.-C.}\ \bibnamefont {Zou}}, \bibinfo {author} {\bibfnamefont {H.-F.}\ \bibnamefont {Zhang}}, \bibinfo {author} {\bibfnamefont {W.-C.}\ \bibnamefont {Xiao}}, \bibinfo {author} {\bibfnamefont {Y.-H.}\ \bibnamefont {Huang}}, \ and\ \bibinfo {author} {\bibfnamefont {X.-D.}\ \bibnamefont {Wang}},\ }\bibfield  {title} {\enquote {\bibinfo {title} {Simulation and experimental comparison of the performance of four-corner-readout plastic scintillator muon-detector system},}\ }\href {\doibase 10.1007/s41365-024-01530-1} {\bibfield  {journal} {\bibinfo  {journal} {Nuclear Science and Techniques}\ }\textbf {\bibinfo {volume} {35}} (\bibinfo {year} {2024}),\ 10.1007/s41365-024-01530-1}\BibitemShut {NoStop}%
\bibitem [{\citenamefont {Valencia}\ \emph {et~al.}(2024)\citenamefont {Valencia}, \citenamefont {Hecht}, \citenamefont {Morris}, \citenamefont {Guardincerri}, \citenamefont {Poulson}, \citenamefont {Bacon},\ and\ \citenamefont {Durham}}]{Valencia:2024dtn}%
  \BibitemOpen
  \bibfield  {author} {\bibinfo {author} {\bibfnamefont {J.~J.}\ \bibnamefont {Valencia}}, \bibinfo {author} {\bibfnamefont {A.~A.}\ \bibnamefont {Hecht}}, \bibinfo {author} {\bibfnamefont {C.~L.}\ \bibnamefont {Morris}}, \bibinfo {author} {\bibfnamefont {E.}~\bibnamefont {Guardincerri}}, \bibinfo {author} {\bibfnamefont {D.}~\bibnamefont {Poulson}}, \bibinfo {author} {\bibfnamefont {J.}~\bibnamefont {Bacon}}, \ and\ \bibinfo {author} {\bibfnamefont {J.~M.}\ \bibnamefont {Durham}},\ }\bibfield  {title} {\enquote {\bibinfo {title} {{First experimental study of multiple orientation muon tomography, with image optimization in sparse data environments}},}\ }\href@noop {} {\  (\bibinfo {year} {2024})},\ \Eprint {http://arxiv.org/abs/2410.07264} {arXiv:2410.07264} \BibitemShut {NoStop}%
\bibitem [{ice(2019)}]{ice2019}%
  \BibitemOpen
  \href {\doibase 10.48550/arXiv.1909.01406} {\enquote {\bibinfo {title} {Seasonal variation of atmospheric muons in {IceCube}},}\ } (\bibinfo {year} {2019}),\ \bibinfo {note} {arXiv:1909.01406 [astro-ph]}\BibitemShut {NoStop}%
\bibitem [{\citenamefont {Tanaka}(2022)}]{Tanaka2022d}%
  \BibitemOpen
  \bibfield  {author} {\bibinfo {author} {\bibfnamefont {H.~K.~M.}\ \bibnamefont {Tanaka}},\ }\bibfield  {title} {\enquote {\bibinfo {title} {Wireless muometric navigation system},}\ }\href {\doibase 10.1038/s41598-022-13280-4} {\bibfield  {journal} {\bibinfo  {journal} {Scientific Reports}\ }\textbf {\bibinfo {volume} {12}},\ \bibinfo {pages} {10114} (\bibinfo {year} {2022})}\BibitemShut {NoStop}%
\bibitem [{\citenamefont {Yang}\ \emph {et~al.}(2022)\citenamefont {Yang}, \citenamefont {Luo}, \citenamefont {Yu}, \citenamefont {Zhao}, \citenamefont {Hu}, \citenamefont {Huang}, \citenamefont {Shen}, \citenamefont {Yang}, \citenamefont {Chen},\ and\ \citenamefont {Tang}}]{yang2022}%
  \BibitemOpen
  \bibfield  {author} {\bibinfo {author} {\bibfnamefont {H.}~\bibnamefont {Yang}}, \bibinfo {author} {\bibfnamefont {G.}~\bibnamefont {Luo}}, \bibinfo {author} {\bibfnamefont {T.}~\bibnamefont {Yu}}, \bibinfo {author} {\bibfnamefont {S.}~\bibnamefont {Zhao}}, \bibinfo {author} {\bibfnamefont {B.}~\bibnamefont {Hu}}, \bibinfo {author} {\bibfnamefont {Z.}~\bibnamefont {Huang}}, \bibinfo {author} {\bibfnamefont {H.}~\bibnamefont {Shen}}, \bibinfo {author} {\bibfnamefont {L.}~\bibnamefont {Yang}}, \bibinfo {author} {\bibfnamefont {Y.}~\bibnamefont {Chen}}, \ and\ \bibinfo {author} {\bibfnamefont {J.}~\bibnamefont {Tang}},\ }\bibfield  {title} {\enquote {\bibinfo {title} {Mugrid: A scintillator detector towards cosmic muon absorption imaging},}\ }\href {\doibase https://doi.org/10.1016/j.nima.2022.167402} {\bibfield  {journal} {\bibinfo  {journal} {Nuclear Instruments and Methods in Physics Research Section A: Accelerators, Spectrometers, Detectors and Associated Equipment}\ }\textbf {\bibinfo {volume} {1042}},\
  \bibinfo {pages} {167402} (\bibinfo {year} {2022})}\BibitemShut {NoStop}%
\bibitem [{sp1(2025)}]{sp101}%
  \BibitemOpen
  \href@noop {} {\enquote {\bibinfo {title} {Beijing hoton official website},}\ }\bibinfo {howpublished} {\url{http://www.hoton.com.cn/yjs/29.html}} (\bibinfo {year} {2025}),\ \bibinfo {note} {accessed: 2025-02-19}\BibitemShut {NoStop}%
\bibitem [{kur(2025)}]{kuraraypsfY11}%
  \BibitemOpen
  \href@noop {} {\enquote {\bibinfo {title} {Wavelength shifting fibers},}\ }\bibinfo {howpublished} {\url{http://kuraraypsf.jp/psf/ws.html}} (\bibinfo {year} {2025}),\ \bibinfo {note} {accessed: 2025-02-19}\BibitemShut {NoStop}%
\bibitem [{ham(2025)}]{hamamatsus13360}%
  \BibitemOpen
  \href@noop {} {\enquote {\bibinfo {title} {Hamamatsu official website},}\ }\bibinfo {howpublished} {\url{https://www.hamamatsu.com/eu/en/product/optical-sensors/mppc/mppc_mppc-array/S13360-1350PE.html}} (\bibinfo {year} {2025}),\ \bibinfo {note} {accessed: 2025-02-19}\BibitemShut {NoStop}%
\bibitem [{\citenamefont {Piemonte}\ and\ \citenamefont {Gola}(2019)}]{Piemonte2019}%
  \BibitemOpen
  \bibfield  {author} {\bibinfo {author} {\bibfnamefont {C.}~\bibnamefont {Piemonte}}\ and\ \bibinfo {author} {\bibfnamefont {A.}~\bibnamefont {Gola}},\ }\bibfield  {title} {\enquote {\bibinfo {title} {Overview on the main parameters and technology of modern {Silicon} {Photomultipliers}},}\ }\href {\doibase 10.1016/j.nima.2018.11.119} {\bibfield  {journal} {\bibinfo  {journal} {Nuclear Instruments and Methods in Physics Research Section A: Accelerators, Spectrometers, Detectors and Associated Equipment}\ }\bibinfo {series} {Silicon {Photomultipliers}: {Technology}, {Characterisation} and {Applications}},\ \textbf {\bibinfo {volume} {926}},\ \bibinfo {pages} {2--15} (\bibinfo {year} {2019})}\BibitemShut {NoStop}%
\bibitem [{\citenamefont {Francesco}\ \emph {et~al.}(2016)\citenamefont {Francesco}, \citenamefont {Bugalho}, \citenamefont {Oliveira}, \citenamefont {Pacher}, \citenamefont {Rivetti}, \citenamefont {Rolo}, \citenamefont {Silva}, \citenamefont {Silva},\ and\ \citenamefont {Varela}}]{Francesco2016}%
  \BibitemOpen
  \bibfield  {author} {\bibinfo {author} {\bibfnamefont {A.~D.}\ \bibnamefont {Francesco}}, \bibinfo {author} {\bibfnamefont {R.}~\bibnamefont {Bugalho}}, \bibinfo {author} {\bibfnamefont {L.}~\bibnamefont {Oliveira}}, \bibinfo {author} {\bibfnamefont {L.}~\bibnamefont {Pacher}}, \bibinfo {author} {\bibfnamefont {A.}~\bibnamefont {Rivetti}}, \bibinfo {author} {\bibfnamefont {M.}~\bibnamefont {Rolo}}, \bibinfo {author} {\bibfnamefont {J.}~\bibnamefont {Silva}}, \bibinfo {author} {\bibfnamefont {R.}~\bibnamefont {Silva}}, \ and\ \bibinfo {author} {\bibfnamefont {J.}~\bibnamefont {Varela}},\ }\bibfield  {title} {\enquote {\bibinfo {title} {Tofpet2: a high-performance asic for time and amplitude measurements of sipm signals in time-of-flight applications},}\ }\href {\doibase 10.1088/1748-0221/11/03/C03042} {\bibfield  {journal} {\bibinfo  {journal} {Journal of Instrumentation}\ }\textbf {\bibinfo {volume} {11}},\ \bibinfo {pages} {C03042} (\bibinfo {year} {2016})}\BibitemShut {NoStop}%
\bibitem [{\citenamefont {Bugalho}\ \emph {et~al.}(2019)\citenamefont {Bugalho}, \citenamefont {Francesco}, \citenamefont {Ferramacho}, \citenamefont {Leong}, \citenamefont {Niknejad}, \citenamefont {Oliveira}, \citenamefont {Rolo}, \citenamefont {Silva}, \citenamefont {Silva}, \citenamefont {Silveira}, \citenamefont {Tavernier},\ and\ \citenamefont {Varela}}]{Bugalho2019}%
  \BibitemOpen
  \bibfield  {author} {\bibinfo {author} {\bibfnamefont {R.}~\bibnamefont {Bugalho}}, \bibinfo {author} {\bibfnamefont {A.~D.}\ \bibnamefont {Francesco}}, \bibinfo {author} {\bibfnamefont {L.}~\bibnamefont {Ferramacho}}, \bibinfo {author} {\bibfnamefont {C.}~\bibnamefont {Leong}}, \bibinfo {author} {\bibfnamefont {T.}~\bibnamefont {Niknejad}}, \bibinfo {author} {\bibfnamefont {L.}~\bibnamefont {Oliveira}}, \bibinfo {author} {\bibfnamefont {M.}~\bibnamefont {Rolo}}, \bibinfo {author} {\bibfnamefont {J.}~\bibnamefont {Silva}}, \bibinfo {author} {\bibfnamefont {R.}~\bibnamefont {Silva}}, \bibinfo {author} {\bibfnamefont {M.}~\bibnamefont {Silveira}}, \bibinfo {author} {\bibfnamefont {S.}~\bibnamefont {Tavernier}}, \ and\ \bibinfo {author} {\bibfnamefont {J.}~\bibnamefont {Varela}},\ }\bibfield  {title} {\enquote {\bibinfo {title} {Experimental characterization of the tofpet2 asic},}\ }\href {\doibase 10.1088/1748-0221/14/03/P03029} {\bibfield  {journal} {\bibinfo  {journal} {Journal of Instrumentation}\ }\textbf
  {\bibinfo {volume} {14}},\ \bibinfo {pages} {P03029} (\bibinfo {year} {2019})}\BibitemShut {NoStop}%
\bibitem [{\citenamefont {Grupen}\ and\ \citenamefont {Shwartz}(2008)}]{Grupen2023}%
  \BibitemOpen
  \bibfield  {author} {\bibinfo {author} {\bibfnamefont {C.}~\bibnamefont {Grupen}}\ and\ \bibinfo {author} {\bibfnamefont {B.}~\bibnamefont {Shwartz}},\ }\href {\doibase 10.1017/9781009401531} {\emph {\bibinfo {title} {Particle Detectors}}}\ (\bibinfo  {publisher} {Cambridge University Press},\ \bibinfo {year} {2008})\ Chap.\ \bibinfo {chapter} {2.5}, pp.\ \bibinfo {pages} {65--67}\BibitemShut {NoStop}%
\bibitem [{\citenamefont {Domingo-Pardo}\ \emph {et~al.}(2009)\citenamefont {Domingo-Pardo}, \citenamefont {Goel}, \citenamefont {Engert}, \citenamefont {Gerl}, \citenamefont {Isaka}, \citenamefont {Kojouharov},\ and\ \citenamefont {Schaffner}}]{DomingoPardo2009}%
  \BibitemOpen
  \bibfield  {author} {\bibinfo {author} {\bibfnamefont {C.}~\bibnamefont {Domingo-Pardo}}, \bibinfo {author} {\bibfnamefont {N.}~\bibnamefont {Goel}}, \bibinfo {author} {\bibfnamefont {T.}~\bibnamefont {Engert}}, \bibinfo {author} {\bibfnamefont {J.}~\bibnamefont {Gerl}}, \bibinfo {author} {\bibfnamefont {M.}~\bibnamefont {Isaka}}, \bibinfo {author} {\bibfnamefont {I.}~\bibnamefont {Kojouharov}}, \ and\ \bibinfo {author} {\bibfnamefont {H.}~\bibnamefont {Schaffner}},\ }\bibfield  {title} {\enquote {\bibinfo {title} {A position sensitive $\gamma $-ray scintillator detector with enhanced spatial resolution, linearity, and field of view},}\ }\href {\doibase 10.1109/tmi.2009.2027436} {\bibfield  {journal} {\bibinfo  {journal} {IEEE Transactions on Medical Imaging}\ }\textbf {\bibinfo {volume} {28}},\ \bibinfo {pages} {2007–2014} (\bibinfo {year} {2009})}\BibitemShut {NoStop}%
\bibitem [{\citenamefont {Cossio}\ \emph {et~al.}(2018)\citenamefont {Cossio}, \citenamefont {Alexeev}, \citenamefont {Amoroso} \emph {et~al.}}]{Cossio2018}%
  \BibitemOpen
  \bibfield  {author} {\bibinfo {author} {\bibfnamefont {F.}~\bibnamefont {Cossio}}, \bibinfo {author} {\bibfnamefont {M.}~\bibnamefont {Alexeev}}, \bibinfo {author} {\bibfnamefont {A.}~\bibnamefont {Amoroso}},  \emph {et~al.},\ }\bibfield  {title} {\enquote {\bibinfo {title} {Optimization of the reconstruction algorithm in triple-{GEM} detector},}\ }in\ \href {\doibase 10.1109/NSSMIC.2018.8824331} {\emph {\bibinfo {booktitle} {2018 {IEEE} {Nuclear} {Science} {Symposium} and {Medical} {Imaging} {Conference} {Proceedings} ({NSS}/{MIC})}}}\ (\bibinfo {year} {2018})\ pp.\ \bibinfo {pages} {1--5},\ \bibinfo {note} {iSSN: 2577-0829}\BibitemShut {NoStop}%
\bibitem [{\citenamefont {Niess}\ \emph {et~al.}(2019)\citenamefont {Niess}, \citenamefont {Barnoud}, \citenamefont {C\^arloganu},\ and\ \citenamefont {Martineau-Huynh}}]{Niess:2019hdn}%
  \BibitemOpen
  \bibfield  {author} {\bibinfo {author} {\bibfnamefont {V.}~\bibnamefont {Niess}}, \bibinfo {author} {\bibfnamefont {A.}~\bibnamefont {Barnoud}}, \bibinfo {author} {\bibfnamefont {C.}~\bibnamefont {C\^arloganu}}, \ and\ \bibinfo {author} {\bibfnamefont {O.}~\bibnamefont {Martineau-Huynh}},\ }\bibfield  {title} {\enquote {\bibinfo {title} {{TURTLE: A C library for an optimistic stepping through a topography}},}\ }\href {\doibase 10.1016/j.cpc.2019.106952} {\  (\bibinfo {year} {2019}),\ 10.1016/j.cpc.2019.106952},\ \Eprint {http://arxiv.org/abs/1904.03435} {arXiv:1904.03435 [physics.comp-ph]} \BibitemShut {NoStop}%
\bibitem [{\citenamefont {Zurlo}\ \emph {et~al.}(2022)\citenamefont {Zurlo}, \citenamefont {Bonomi}, \citenamefont {Donzella}, \citenamefont {Pagano}, \citenamefont {Zenoni},\ and\ \citenamefont {Zumerle}}]{Zurlo:2022ltm}%
  \BibitemOpen
  \bibfield  {author} {\bibinfo {author} {\bibfnamefont {N.}~\bibnamefont {Zurlo}}, \bibinfo {author} {\bibfnamefont {G.}~\bibnamefont {Bonomi}}, \bibinfo {author} {\bibfnamefont {A.}~\bibnamefont {Donzella}}, \bibinfo {author} {\bibfnamefont {D.}~\bibnamefont {Pagano}}, \bibinfo {author} {\bibfnamefont {A.}~\bibnamefont {Zenoni}}, \ and\ \bibinfo {author} {\bibfnamefont {G.}~\bibnamefont {Zumerle}},\ }\bibfield  {title} {\enquote {\bibinfo {title} {{A new Monte Carlo muon generator for cosmic-ray muon applications}},}\ }\href {\doibase 10.22323/1.409.0019} {\bibfield  {journal} {\bibinfo  {journal} {PoS}\ }\textbf {\bibinfo {volume} {CompTools2021}},\ \bibinfo {pages} {019} (\bibinfo {year} {2022})}\BibitemShut {NoStop}%
\bibitem [{\citenamefont {Niess}(2022)}]{Niess2022}%
  \BibitemOpen
  \bibfield  {author} {\bibinfo {author} {\bibfnamefont {V.}~\bibnamefont {Niess}},\ }\bibfield  {title} {\enquote {\bibinfo {title} {{The PUMAS library}},}\ }\href {\doibase 10.1016/j.cpc.2022.108438} {\bibfield  {journal} {\bibinfo  {journal} {Comput. Phys. Commun.}\ }\textbf {\bibinfo {volume} {279}},\ \bibinfo {pages} {108438} (\bibinfo {year} {2022})},\ \Eprint {http://arxiv.org/abs/2206.01457} {arXiv:2206.01457 [physics.comp-ph]} \BibitemShut {NoStop}%
\bibitem [{\citenamefont {Agostinelli}\ \emph {et~al.}(2003)\citenamefont {Agostinelli}, \citenamefont {Allison}, \citenamefont {Amako} \emph {et~al.}}]{Agostinelli2003}%
  \BibitemOpen
  \bibfield  {author} {\bibinfo {author} {\bibfnamefont {S.}~\bibnamefont {Agostinelli}}, \bibinfo {author} {\bibfnamefont {J.}~\bibnamefont {Allison}}, \bibinfo {author} {\bibfnamefont {K.}~\bibnamefont {Amako}},  \emph {et~al.},\ }\bibfield  {title} {\enquote {\bibinfo {title} {Geant4—a simulation toolkit},}\ }\href {\doibase 10.1016/S0168-9002(03)01368-8} {\bibfield  {journal} {\bibinfo  {journal} {Nuclear Instruments and Methods in Physics Research Section A: Accelerators, Spectrometers, Detectors and Associated Equipment}\ }\textbf {\bibinfo {volume} {506}},\ \bibinfo {pages} {250--303} (\bibinfo {year} {2003})}\BibitemShut {NoStop}%
\bibitem [{\citenamefont {Das}\ \emph {et~al.}(2022)\citenamefont {Das}, \citenamefont {Tripathy}, \citenamefont {Jagga}, \citenamefont {Bhattacharya}, \citenamefont {Majumdar},\ and\ \citenamefont {Mukhopadhyay}}]{instruments6040077}%
  \BibitemOpen
  \bibfield  {author} {\bibinfo {author} {\bibfnamefont {S.}~\bibnamefont {Das}}, \bibinfo {author} {\bibfnamefont {S.}~\bibnamefont {Tripathy}}, \bibinfo {author} {\bibfnamefont {P.}~\bibnamefont {Jagga}}, \bibinfo {author} {\bibfnamefont {P.}~\bibnamefont {Bhattacharya}}, \bibinfo {author} {\bibfnamefont {N.}~\bibnamefont {Majumdar}}, \ and\ \bibinfo {author} {\bibfnamefont {S.}~\bibnamefont {Mukhopadhyay}},\ }\bibfield  {title} {\enquote {\bibinfo {title} {Muography for inspection of civil structures},}\ }\href {\doibase 10.3390/instruments6040077} {\bibfield  {journal} {\bibinfo  {journal} {Instruments}\ }\textbf {\bibinfo {volume} {6}} (\bibinfo {year} {2022}),\ 10.3390/instruments6040077}\BibitemShut {NoStop}%
\end{thebibliography}%

\end{document}